\documentclass[aps,prc,superscriptaddress,showpacs,floatfix,letterpaper,preprintnumbers,nofootinbib,showkeys]{revtex4-2}
\usepackage{epsfig}
\usepackage{bm}
\usepackage{amsmath}
\usepackage{amstext}
\usepackage{amssymb}
\usepackage{mathrsfs}
\usepackage{color}
\usepackage{subfigure}
\usepackage{graphicx}

\usepackage{hyperref}
\hypersetup{
    colorlinks=true,
    linkcolor=blue,
    filecolor=magenta,      
    urlcolor=blue,
    citecolor=blue
}

\newcommand{\trento}{\texttt{T$_R$ENTo} }

\begin{document}

\title{Multi-scale evolution of charmed particles in a nuclear medium}

\author{W.~Fan}
\email[Corresponding author: ]{wenkai.fan@duke.edu}
\affiliation{Department of Physics, Duke University, Durham NC 27708.}

\author{G.~Vujanovic}
\email[Corresponding author: ]{gojko.vujanovic@uregina.ca}
\affiliation{Department of Physics and Astronomy, Wayne State University, Detroit MI 48201.}
\affiliation{Department of Physics, University of Regina, Regina, SK S4S 0A2, Canada.}

\author{S.~A.~Bass}
\affiliation{Department of Physics, Duke University, Durham NC 27708.}

\author{A.~Majumder}
\affiliation{Department of Physics and Astronomy, Wayne State University, Detroit MI 48201.}




\author{A.~Angerami}
\affiliation{Lawrence Livermore National Laboratory, Livermore CA 94550.}

\author{R.~Arora}
\affiliation{Research Computing Group, University Technology Solutions, The University of Texas at San Antonio, San Antonio TX 78249.}

\author{S.~Cao}
\affiliation{Institute of Frontier and Interdisciplinary Science, Shandong University, Qingdao, Shandong 266237, China}
\affiliation{Department of Physics and Astronomy, Wayne State University, Detroit MI 48201.}

\author{Y.~Chen}
\affiliation{Laboratory for Nuclear Science, Massachusetts Institute of Technology, Cambridge MA 02139.}
\affiliation{Department of Physics, Massachusetts Institute of Technology, Cambridge MA 02139.}

\author{T.~Dai}
\affiliation{Department of Physics, Duke University, Durham NC 27708.}

\author{L.~Du}
\affiliation{Department of Physics, McGill University, Montr\'{e}al QC H3A\,2T8, Canada.}

\author{R.~Ehlers}
\affiliation{Department of Physics and Astronomy, University of Tennessee, Knoxville TN 37996.}
\affiliation{Physics Division, Oak Ridge National Laboratory, Oak Ridge TN 37830.}

\author{H.~Elfner}
\affiliation{GSI Helmholtzzentrum f\"{u}r Schwerionenforschung, 64291 Darmstadt, Germany.}
\affiliation{Institute for Theoretical Physics, Goethe University, 60438 Frankfurt am Main, Germany.}
\affiliation{Frankfurt Institute for Advanced Studies, 60438 Frankfurt am Main, Germany.}

\author{R.~J.~Fries}
\affiliation{Cyclotron Institute, Texas A\&M University, College Station TX 77843.}
\affiliation{Department of Physics and Astronomy, Texas A\&M University, College Station TX 77843.}

\author{C.~Gale}
\affiliation{Department of Physics, McGill University, Montr\'{e}al QC H3A\,2T8, Canada.}


\author{Y.~He}
\affiliation{Guangdong Provincial Key Laboratory of Nuclear Science, Institute of Quantum Matter, South China Normal University, Guangzhou 510006, China.}
\affiliation{Guangdong-Hong Kong Joint Laboratory of Quantum Matter, Southern Nuclear Science Computing Center, South China Normal University, Guangzhou 510006, China.}

\author{M.~Heffernan}
\affiliation{Department of Physics, McGill University, Montr\'{e}al QC H3A\,2T8, Canada.}

\author{U.~Heinz}
\affiliation{Department of Physics, The Ohio State University, Columbus OH 43210.}

\author{B.~V.~Jacak}
\affiliation{Department of Physics, University of California, Berkeley CA 94270.}
\affiliation{Nuclear Science Division, Lawrence Berkeley National Laboratory, Berkeley CA 94270.}

\author{P.~M.~Jacobs}
\affiliation{Department of Physics, University of California, Berkeley CA 94270.}
\affiliation{Nuclear Science Division, Lawrence Berkeley National Laboratory, Berkeley CA 94270.}

\author{S.~Jeon}
\affiliation{Department of Physics, McGill University, Montr\'{e}al QC H3A\,2T8, Canada.}

\author{Y.~Ji}
\affiliation{Department of Statistical Science, Duke University, Durham NC 27708.}

\author{K.~Kauder}
\affiliation{Department of Physics, Brookhaven National Laboratory, Upton NY 11973.}

\author{L.~Kasper}
\affiliation{Department of Physics and Astronomy, Vanderbilt University, Nashville TN 37235.}

\author{W.~Ke}
\affiliation{Los Alamos National Laboratory, Theoretical Division, Los Alamos, NM 87545.}

\author{M.~Kelsey}
\affiliation{Department of Physics and Astronomy, Wayne State University, Detroit MI 48201.}


\author{M.~Kordell~II}
\affiliation{Cyclotron Institute, Texas A\&M University, College Station TX 77843.}
\affiliation{Department of Physics and Astronomy, Texas A\&M University, College Station TX 77843.}

\author{A.~Kumar}
\affiliation{Department of Physics, McGill University, Montr\'{e}al QC H3A\,2T8, Canada.}
\affiliation{Department of Physics and Astronomy, Wayne State University, Detroit MI 48201.}

\author{J.~Latessa}
\affiliation{Department of Computer Science, Wayne State University, Detroit MI 48202.}

\author{Y.-J.~Lee}
\affiliation{Laboratory for Nuclear Science, Massachusetts Institute of Technology, Cambridge MA 02139.}
\affiliation{Department of Physics, Massachusetts Institute of Technology, Cambridge MA 02139.}

\author{D.~Liyanage}
\affiliation{Department of Physics, The Ohio State University, Columbus OH 43210.}

\author{A.~Lopez}
\affiliation{Instituto  de  F\`{i}sica,  Universidade  de  S\~{a}o  Paulo,  C.P.  66318,  05315-970  S\~{a}o  Paulo,  SP,  Brazil. }

\author{M.~Luzum}
\affiliation{Instituto  de  F\`{i}sica,  Universidade  de  S\~{a}o  Paulo,  C.P.  66318,  05315-970  S\~{a}o  Paulo,  SP,  Brazil. }

\author{S.~Mak}
\affiliation{Department of Statistical Science, Duke University, Durham NC 27708.}

\author{A.~Mankolli}
\affiliation{Department of Physics and Astronomy, Vanderbilt University, Nashville TN 37235.}

\author{C.~Martin}
\affiliation{Department of Physics and Astronomy, University of Tennessee, Knoxville TN 37996.}

\author{H.~Mehryar}
\affiliation{Department of Computer Science, Wayne State University, Detroit MI 48202.}

\author{T.~Mengel}
\affiliation{Department of Physics and Astronomy, University of Tennessee, Knoxville TN 37996.}

\author{J.~Mulligan}
\affiliation{Department of Physics, University of California, Berkeley CA 94270.}
\affiliation{Nuclear Science Division, Lawrence Berkeley National Laboratory, Berkeley CA 94270.}

\author{C.~Nattrass}
\affiliation{Department of Physics and Astronomy, University of Tennessee, Knoxville TN 37996.}

\author{D.~Oliinychenko}
\affiliation{Nuclear Science Division, Lawrence Berkeley National Laboratory, Berkeley CA 94270.}
\affiliation{Institute for Nuclear Theory, University of Washington, Seattle WA, 98195.}

\author{J.-F. Paquet}
\affiliation{Department of Physics, Duke University, Durham NC 27708.}

\author{J.~H.~Putschke}
\affiliation{Department of Physics and Astronomy, Wayne State University, Detroit MI 48201.}

\author{G.~Roland}
\affiliation{Laboratory for Nuclear Science, Massachusetts Institute of Technology, Cambridge MA 02139.}
\affiliation{Department of Physics, Massachusetts Institute of Technology, Cambridge MA 02139.}

\author{B.~Schenke}
\affiliation{Physics Department, Brookhaven National Laboratory, Upton NY 11973.}

\author{L.~Schwiebert}
\affiliation{Department of Computer Science, Wayne State University, Detroit MI 48202.}

\author{A.~Sengupta}
\affiliation{Cyclotron Institute, Texas A\&M University, College Station TX 77843.}
\affiliation{Department of Physics and Astronomy, Texas A\&M University, College Station TX 77843.}

\author{C.~Shen}
\affiliation{Department of Physics and Astronomy, Wayne State University, Detroit MI 48201.}
\affiliation{RIKEN BNL Research Center, Brookhaven National Laboratory, Upton NY 11973.}

\author{A.~Silva}
\affiliation{Department of Physics and Astronomy, University of Tennessee, Knoxville TN 37996.}

\author{C.~Sirimanna}
\affiliation{Department of Physics and Astronomy, Wayne State University, Detroit MI 48201.}

\author{D.~Soeder}
\affiliation{Department of Physics, Duke University, Durham NC 27708.}

\author{R.~A.~Soltz}
\affiliation{Department of Physics and Astronomy, Wayne State University, Detroit MI 48201.}
\affiliation{Lawrence Livermore National Laboratory, Livermore CA 94550.}

\author{I.~Soudi}
\affiliation{Department of Physics and Astronomy, Wayne State University, Detroit MI 48201.}

\author{J.~Staudenmaier}
\affiliation{Institute for Theoretical Physics, Goethe University, 60438 Frankfurt am Main, Germany.}

\author{M.~Strickland}
\affiliation{Department of Physics, Kent State University, Kent, OH 44242.}

\author{Y.~Tachibana}
\affiliation{Akita International University, Yuwa, Akita-city 010-1292, Japan.}
\affiliation{Department of Physics and Astronomy, Wayne State University, Detroit MI 48201.}

\author{J.~Velkovska}
\affiliation{Department of Physics and Astronomy, Vanderbilt University, Nashville TN 37235.}

\author{X.-N.~Wang}
\affiliation{Key Laboratory of Quark and Lepton Physics (MOE) and Institute of Particle Physics, Central China Normal University, Wuhan 430079, China.}
\affiliation{Department of Physics, University of California, Berkeley CA 94270.}
\affiliation{Nuclear Science Division, Lawrence Berkeley National Laboratory, Berkeley CA 94270.}

\author{R.~L.~Wolpert}
\affiliation{Department of Statistical Science, Duke University, Durham NC 27708.}

\author{W.~Zhao}
\affiliation{Department of Physics and Astronomy, Wayne State University, Detroit MI 48201.}

\collaboration{The JETSCAPE Collaboration}

\begin{abstract}
Parton energy-momentum exchange with the quark gluon plasma (QGP) is a multi-scale problem. In this work, we calculate the interaction of charm quarks with the QGP within the higher twist formalism at high virtuality and high energy using the MATTER model, while the low virtuality and high energy portion is treated via a Linearized Boltzmann Transport (LBT) formalism. Coherence effect that reduces the medium-induced emission rate in the MATTER model is also taken into account through a virtuality-dependent $\hat{q}$, leaving the simultaneous dependence of $\hat{q}$ on heavy quark mass and virtuality for future studies. The interplay between these two formalisms is studied phenomenologically and used to produce a first description of the D-meson and charged hadron nuclear modification factor $R_{AA}$ across multiple centralities. All calculations were carried out utilizing the JETSCAPE framework.
\end{abstract}

\date{\today}
\maketitle
\section{Introduction}
Understanding the properties of the quark gluon plasma (QGP) at various energy scales is at the core of the ongoing relativistic heavy-ion program at the Relativistic Heavy-Ion Collider (RHIC) and the Large Hadron Collider (LHC) \cite{Wang:2016opj}. Soft hadronic observables (e.g. the multiplicity of various hadronic species or their anisotropic flow) probe transport properties of the QGP such as its shear and bulk viscosity and can thus be used to constrain the latter \cite{Bernhard:2019bmu}. On the other hand, penetrating probes such as electromagnetic radiation (see e.g. \cite{Vujanovic:2019yih,Tripolt:2020dac,Geurts:2020ogz} and references therein) or high energy phenomena, such as jets and heavy flavor production, give access to properties of the medium at high temperatures (see e.g. \cite{Cao:2020wlm,Connors:2017ptx,Djordjevic:2008iz} and reference therein). Jets and heavy flavor production in the vacuum are well understood and calculable using perturbative QCD techniques \cite{Frawley:2008kk} as well as Monte Carlo generators of parton showers such as PYTHIA \cite{Sjostrand:2019zhc, JETSCAPE:2019udz}. This provides a baseline against which the nuclear medium modifications of these quantities give insight into the QGP properties.

Medium-induced modifications of parton showers are encapsulated in QGP transport coefficients \cite{Nahrgang:2016lst,Brambilla:2020siz,Kumar:2020wvb}. Transverse momentum broadening ($\hat{q}$) of parton showers in the QGP is a notable example of such medium-induced interactions. More formally, 
\begin{equation}
\hat{q}=\frac{\langle p^2_T\rangle_L}{L},
\label{eq:qhat}
\end{equation}
where $\langle p^2_T\rangle_L$ corresponds to the squared transverse momentum change of a parton as it traverses a distance $L$ through the QGP medium before splitting, and thus $\hat{q}$ is the average transverse momentum change per unit length. 

As parton interactions change at different energy and virtuality scales, a framework that allows for a systematic investigation of their medium-induced interactions is needed \cite{Putschke:2019yrg}, such as that provided by the Jet Energy-loss Tomography with a Statistically and Computationally Advanced Program Envelope (JETSCAPE) Collaboration. The holistic approach taken by JETSCAPE has improved both, our understanding of the bulk transport coefficients of the QGP, such as the shear and bulk viscosity \cite{Paquet:2020rxl, JETSCAPE:2020shq,JETSCAPE:2020mzn}, as well as the jet energy-loss transport coefficient $\hat{q}$ \cite{JETSCAPE:2017eso,JETSCAPE:2021ehl}. Our goal in this work is to describe the evolution of heavy quarks, specifically charm quarks, within the QGP using the JETSCAPE framework, and to explore how the multi-scale physics included in the JETSCAPE framework affect the leading D-meson as well as the charged hadron nuclear modification factor $R_{AA}$. While, JETSCAPE version 3.1 is used in our simulation, the current public version (i.e. v3.5) of JETSCAPE \cite{JETSCAPE3.5} contains the same physics as explored herein and can be used instead.

The multi-scale problem of parton interactions with the QGP, often called parton ``energy loss'' for brevity, can be loosely separated into three regimes: one of high energy ($E$) and high virtuality ($t$), followed by a high-$E$ and low-$t$ regime both described via perturbation theory, ultimately ending up in the low-$E$ and low-$t$ phase space where non-perturbative phenomena take place.~\footnote{Past efforts \cite{Burke:2013yra} focused more on developing the theoretical formalisms and/or numerical approaches to describe these kinematic regimes, and thus often a single formalism was used throughout the entire evolution history of the parton shower. The JETSCAPE framework provides the opportunity to investigate multiple regimes in a consistent fashion.} All three sectors are incorporated inside the JETSCAPE framework, with the first two being the focus of this study. 

Starting in the high-$E$ and high-$t$ region of phase space, any virtual particle will undergo multiple radiations and thereby reduce its virtuality. Such processes are already described in the vacuum using Monte Carlo shower generators, such as PYTHIA, which we here use solely to sample the hard process giving rise to a parton shower. The subsequent virtuality-ordered evolution of the shower profile, both in position and momentum spaces, will be simulated using medium-modified interactions between hard partons and the QGP following the in-medium Dokshitzer-Gribov-Lipatov-Altarelli-Parisi (DGLAP) evolution \cite{Majumder:2009zu,Wang:2009qb} based on the higher-twist formalism \cite{Wang:2001ifa,Majumder:2009ge,Qin:2009gw} valid when $t^2\gg \hat{q}E$. The medium-modified DGLAP evolution is typically stopped once the virtuality reaches $t\sim\sqrt{\hat{q} E}$.

When the virtuality scale is close to that of the medium, rate equations \cite{Jeon:2003gi,Turbide:2005fk,Qin:2009bk} become an apt description of parton evolution in the QGP. Formalisms based on Baier-Dokshitzer-Mueller-Peigne-Schiff- (BDMPS) \cite{Baier:1996kr,Baier:1996sk}, including Zakharov's \cite{Zakharov:1996fv} contribution (BDMPS-Z), or the Arnold-Moore-Yaffe (AMY) \cite{Arnold:2001ms,Arnold:2002ja,Arnold:2002zm} approach, the Djordjevic-Gyulassy-Levai-Vitev \cite{Djordjevic:2003zk,Gyulassy:1999zd} prescription, as well as those inspired from higher twist \cite{Wang:2013cia,Cao:2016gvr,Cao:2017hhk,Chen:2017zte} have all be used in the past. In this work, we follow the higher twist-inspired rate equations approach. Finally, once partons reach the low-$E$ low-$t$ region, they will be handed off to PYTHIA for hadronization. 

This work is organized as follows: In Sec. \ref{sec:medium evol} we present details about the hydrodynamical simulation of the QGP through which partons will interact, Sec. \ref{sec:energy_loss} describes the multi-stage energy loss models used in this JETSCAPE calculation, Sec. \ref{sec:results} presents the results of this calculation, while Sec. \ref{sec:conclusion} is reserved for concluding remarks. 

\section{Evolution of the QCD medium}
\label{sec:medium evol}
The evolution of the QCD medium used throughout this study is performed using a Bayesian tuned boost-invariant 2+1-dimensional hydrodynamic-inspired model which involves three stages: a pre-hydrodynamic, hydrodynamic, and a hadronic transport stage~\cite{Bernhard:2019bmu,Shen:2014vra, Bass:1998ca,Bleicher:1999xi}. The pre-hydrodynamic stage is composed of the \trento{} (initial condition for Pb-Pb collisions) \cite{Moreland:2014oya}, followed by free-streaming for a proper time of $\tau_{FS}=1.2$ fm/$c$. This generates a non-trivial initial condition for the hydrodynamical simulation to follow. We have generated in total 400 \trento{} initial Pb-Pb configurations in the 0-10\% centrality class at $\sqrt{s_{NN}}=5.02$ TeV. The relevant parameters used for simulating the evolution of the QCD medium are extracted from a Bayesian model-to-data comparison, explained in \cite{Bernhard:2019bmu,Bernhard:2018hnz}.

The hydrodynamical simulation \cite{Song:2007ux,Shen:2014vra} is performed until the the cross-over temperature of $T_c=154$ MeV is reached~\cite{Bazavov:2014pvz}, at which point fluid fields are converted into particles \cite{Bernhard:2018hnz,Huovinen:2012is} whose subsequent evolution is governed by hadronic Boltzmann transport \cite{Bass:1998ca,Bleicher:1999xi}.

All parotns, including charm quarks, do not interact during the pre-hydrodynamical evolution as it is given by free streaming. Since we shall focus on charm quarks and D-mesons at momenta above $7$~GeV, hadronic final state interactions are negligible as well. Thus, all parton (and charm quarks in particular) only interact during the hydrodynamical portion of the evolution, which is given by second-order Israel-Stewart theory \cite{Israel1976310,Israel:1979wp}. Other than conservation of energy and momentum, second-order hydrodynamical equations also include relaxation-type equation for six independent viscous degrees of freedom, namely five in the shear tensor $\pi^{\mu\nu}$ and one for bulk pressure $\Pi$. The energy-momentum conservation equation reads:
\begin{eqnarray}
\partial_\mu T^{\mu\nu}&=&0,\nonumber\\
T^{\mu\nu}&=& \epsilon u^\mu u^\nu - (P+\Pi)\Delta^{\mu\nu}+\pi^{\mu\nu}
\label{eq:energy-momentum_conserv}
\end{eqnarray}  
where $\epsilon$ is the energy density, $u^\mu$ is the flow four-velocity, $P$ is the thermodynamic pressure related to $\epsilon$ by the equation of state $P(\epsilon)$ \cite{Bazavov:2014pvz,Bernhard:2018hnz}, $\Delta^{\mu\nu}=g^{\mu\nu}-u^\mu u^\nu$ projects on the spatial directions in the local fluid rest frame, and $g^{\mu\nu}={\rm diag}(1,-1,-1,-1)$ is the metric tensor. The dissipative degrees of freedom satisfy:
\begin{eqnarray}
\tau_\Pi \dot{\Pi}+\Pi &=& -\zeta\theta - \delta_{\Pi\Pi}\Pi\theta + \lambda_{\Pi\pi}\pi^{\alpha\beta}\sigma_{\alpha\beta}, \label{eq:bulk_relax}\\
\tau_\pi \dot{\pi}^{\langle\mu\nu\rangle}+\pi^{\mu\nu} &=& 2\eta\sigma^{\mu\nu}-\delta_{\pi\pi}\pi^{\mu\nu}\theta + \lambda_{\pi\Pi}\Pi\sigma^{\mu\nu} - \tau_{\pi\pi} \pi^{\langle\mu}_\alpha\sigma^{\nu\rangle\alpha} + \phi_7 \pi^{\langle\mu}_\alpha\pi^{\nu\rangle\alpha},
\label{eq:shear_relax}
\end{eqnarray}  
where $\dot{\Pi}\equiv u^\alpha\partial_\alpha\Pi$, $\dot{\pi}^{\langle\mu\nu\rangle}\equiv \Delta^{\mu\nu}_{\alpha\beta}u^\lambda\partial_\lambda\pi^{\alpha\beta}$, $\Delta_{\alpha\beta}^{\mu\nu}\equiv\left(\Delta_{\alpha}^{\mu}\Delta_{\beta}^{\nu}+\Delta_{\beta}^{\mu}\Delta_{\alpha}^{\nu}\right)/2-\left(\Delta_{\alpha\beta}\Delta^{\mu\nu}\right)/3$, $\theta\equiv\partial_\alpha u^\alpha$, $\sigma^{\mu\nu}\equiv\partial^{\langle\mu} u^{\nu\rangle}$, with $A^{\langle\mu\nu\rangle}\equiv\Delta^{\mu\nu}_{\alpha\beta}A^{\alpha\beta}$. Other than $\zeta$ and $\eta$, which will be discussed in a moment, the various transport coefficients present in Eqs.~(\ref{eq:bulk_relax}) and (\ref{eq:shear_relax}) were computed assuming a single component gas of constituent particles in the limit $m/T \ll 1$ \cite{Denicol:2012cn,Denicol:2014vaa}, where $m$ is their mass and $T$ the temperature, respectively. These are summarized in Table \ref{table:bulk_shear_relax}, where $c^2_s=\partial P /\partial \epsilon$ is the speed of sound squared. 

\begin{table}[!ht]
\caption{Transport coefficients in Eqs. (\ref{eq:bulk_relax}) and (\ref{eq:shear_relax}).} 
\centering
\begin{tabular}{l l l l l l l l l l}
\hline \hline
Bulk & & $\tau_\Pi=\zeta\left[15(\epsilon+P)\left(\frac{1}{3}-c^2_s\right)^2\right]^{-1}$ & $\delta_{\Pi\Pi}=\frac{2}{3}\tau_\Pi$ & & $\lambda_{\Pi\pi}=\frac{8}{5}\left(\frac{1}{3}-c^2_s\right)\tau_\Pi$ & & \\
Shear & & $\tau_\pi=5\eta\left(\epsilon+P\right)^{-1}$  & $\delta_{\pi\pi}=\frac{4}{3}\tau_\pi$ &  & $\lambda_{\pi\Pi}=\frac{6}{5}\tau_\pi$ & & $\tau_{\pi\pi}=\frac{10}{7}\tau_\pi$ & & $\phi_7=\frac{18}{175} \frac{\tau_\pi}{\eta}$\\
\hline \hline
\end{tabular}
\label{table:bulk_shear_relax} 
\end{table}

The specific shear viscosity ($\eta/s$) --- where $s$ is the entropy density --- and the specific bulk viscosity ($\zeta/s$) are both taken from a recent Bayesian model-to-data comparison~\cite{Bernhard:2019bmu}. 

\section{Parton interactions with the QGP}
\label{sec:energy_loss}
Following initial parton momentum production in PYTHIA and transverse positions sampled from the binary collision profile in \trento, the evolution of high-energy and high-virtuality partons is calculated in Modular All Twist Transverse-scattering Elastic-drag and Radiation (MATTER) \cite{Majumder:2013re} (Modular All Twist Transverse-scattering Elastic-drag and Radiation), which describes their interactions with the QGP using the higher twist formalism \cite{Wang:2001ifa,Majumder:2009ge,Qin:2009gw}. The latter develops a virtuality ordered shower, which this study extends by including heavy quarks in MATTER according to the soft collinear effective theory (SCET) devised by Ref.~\cite{Abir:2015hta}. Once partons in the shower reach the low virtuality (and high energy) regime, further evolution proceeds via the Linearized Boltzmann Transport (LBT) model \cite{Luo:2018pto}. In the LBT formalism, the interactions between the partons and the QGP preserve the virtuality of the partons, while modifying their energy and three-momentum. Partons with virtuality $t > t_s$, $t_s$ being the switching virtuality, are evolved by MATTER, while those with $t\leq t_s$ are evolved using LBT.      

Following the LBT evolution, the JETSCAPE framework determines whether the partons undergo further evolution in MATTER (this can happen if a parton quickly exits the medium and continues to shower in the vacuum, for example) or whether they hadronize (hadronization is handled via fragmentation in PYTHIA).

\subsection{The higher-twist formalism in MATTER }
\label{sec:MATTER}


This section summarizes the physical mechanisms involving heavy flavor. In the higher twist approach, the radiation of a gluon off a heavy quark was first theoretically devised using SCET in Ref. \cite{Abir:2015hta}, while solely the light flavor higher-twist calculations were explored in Refs.~\cite{Wang:2001ifa,Majumder:2009ge,Qin:2009gw,Majumder:2013re}. The radiation process $Q\to Q+g$, where $Q$ is a heavy quark, is \cite{Abir:2015hta}:
\begin{eqnarray}
\frac{dN^{\rm vac}}{dz dt} + \frac{dN^{\rm med}}{dz dt } &=& \frac{\alpha_s(t)}{2\pi} \frac{P_{g\gets Q}(z)}{t}\left\{ 1+ \int^{\tau^+_Q}_0 d\tau^+\frac{1}{z(1-z)t(1+\chi)^2} \left[2-2\cos\left( \frac{\tau^+}{\tau^+_Q} \right)\right]\right.\times\nonumber\\
&\times& \left. \left[  \left(\frac{1+z}{2}\right) - \chi + \left(\frac{1+z}{2}\right) \chi^2 \right]\hat{q}  \right\}
.\nonumber\\
\label{eq:hq_ht}
\end{eqnarray} 
where $z$ is the momentum fraction of the daughter heavy quark, $M$ is the mass of the heavy quark, $\chi=(1-z)^2M^2/l^2_\perp$, with $l^2_\perp$ being the relative transverse momentum square between the outgoing daughter partons, determined via $z(1-z)t=l^2_\perp(1+\chi)$, while $t$ is the virtuality of the heavy quark and $P_{g\gets Q}(z)=C_F\left(\frac{1+z^2}{1-z^{\,\,\,}}\right)$ is the splitting function. The integral over light-cone time $\tau^+$ in Eq. (\ref{eq:hq_ht}) assumes the medium is in its rest frame, with the upper bound $\tau^+_Q=2q^{+}/t$ being given by the ratio of forward light-cone momentum $q^+=\left(q^0+{\bf q}\cdot \hat{n}\right)/\sqrt{2}$ (with $\hat{n}={\bf q}/\vert {\bf q}\vert$), and the virtuality $t$. 

\subsubsection{Transverse momentum broadening of partons in the QGP}
The transverse momentum broadening ($\hat{q}$) acquired by the quark as it traverses the QGP is the only quantity that explicitly depends on $\tau^+$ via its temperature dependence $\hat{q}(T)$. Following the the Hard Thermal Loop (HTL) approximation as presented in Ref. \cite{He:2015pra},  $\hat{q}$ is 
\begin{eqnarray}
\hat{q}^{HTL}=C_a \frac{42\zeta(3)}{\pi}\alpha_s^2T^3 \ln \left(\frac{cET}{4m^2_D}\right)
\end{eqnarray}
where $C_a=N_c=3$ the number of colors, $\zeta(3)\approx 1.20205$ is Ap\'{e}ry's constant, $E$ is the incoming hard parton's energy, while the Debye mass is $m^2_D=\frac{4\pi\alpha_s T^2}{3} \left(N_c+\frac{N_f}{2}\right)$, with the number of flavors $N_f=3$, the temperature $T$, and $c \approx 5.7$ \cite{Caron-Huot:2009fku}. Using a calibration to {\it light flavor} experimental observables, an effective value of $\alpha_s$ namely $\alpha^{(\rm eff)}_s=0.3$ was obtained \cite{JETSCAPE:2022jer}. This formulation of $\hat{q}$ will also be used to study heavy flavor energy loss in this work. The $R_{AA}$ study done in Ref. \cite{JETSCAPE:2022jer} also revealed that a constant effective $\alpha^{(\rm eff)}_s$ can be improved by allowing the coupling to run with the scale $\mu^2=2ET$ via
\begin{eqnarray}
\hat{q}^{HTL}=C_a \frac{42\zeta(3)}{\pi}\alpha_s\left(\mu^2\right) \alpha^{(\rm eff)}_s T^3 \ln \left(\frac{cET}{4m^2_D}\right)
\end{eqnarray}    
where
\begin{eqnarray}
m^2_D&=&\frac{4\pi\alpha^{(\rm eff)}_s T^2}{3} \left(N_c+\frac{N_f}{2}\right)=6\pi\nonumber\alpha^{(\rm eff)}_s T^2\\
\alpha_s\left(\mu^2\right)&=&
\begin{cases}
\alpha^{(\rm eff)}_s & \mu^2 < \mu^2_0, \\
\frac{4\pi}{11-2N_f/3} \frac{1}{\ln\left(\frac{\mu^2}{\Lambda^2}\right)} & \mu^2 > \mu^2_0.
\end{cases}
\end{eqnarray}
with $E$ being the energy of the incoming hard parton participating in a scattering or radiation process, and $\Lambda$ being chosen such that $\alpha_s(\mu^2)=\alpha^{(\rm eff)}_s$ at $\mu^2_0=1$ GeV$^2$ \cite{Kumar:2019uvu}. Thus, in our simulation the incoming hard parton (with energy $E$) has a different coupling (i.e., $\alpha_s(\mu^2)$) than the QGP parton $\left({\rm i.e.,}\,\,\alpha_s^{(\rm eff)}\right)$.

Furthermore, Ref. \cite{Kumar:2019uvu} suggests that $\hat{q}$ changes with the virtuality scale --- beyond the running of $\alpha_s(\mu^2)$ --- which offers an alternative explanation of the puzzle where the extracted $\hat{q}/T^3$ is around 50\% smaller at the LHC compared to the RHIC \cite{Burke:2013yra}.  As the virtuality of the partons increases with energy, the transverse size of the dipole formed by the parton and the emitted gluon decreases, and as a result, can only sample gluons from the medium that have wavelengths comparable to this size. This causes a suppression of $\hat{q}$ with higher parton virtuality. In Ref. \cite{Kumar:2019uvu} an integrated form of $\hat{q}$ was adopted, whereas in this study an effective parametrization of that virtuality dependence \cite{JETSCAPE:2022jer} is used
\begin{eqnarray}
\hat{q}(t) &=& \hat{q}^{HTL}\frac{c_0}{1+c_1\ln^2(t)+c_2\ln^4(t)}\nonumber\\
           &=& C_a \frac{42\zeta(3)}{\pi}\alpha_s\left(\mu^2\right) \alpha^{(\rm eff)}_s T^3 \ln \left(\frac{cET}{4m^2_D}\right)\frac{c_0}{1+c_1\ln^2(t)+c_2\ln^4(t)},
\label{eq:qhat_t}
\end{eqnarray}
where $c_1$ and $c_2$ are tunable parameters, $t$ is the virtuality of the parton, and $c_0=1+c_1\ln^2(t_s)+c_2\ln^4(t_s)$ is an overall normalization ensuring that the $t$-dependent contribution --- given by $\hat{q}(t)/\hat{q}^{HTL}$ ---  is unitless and lies within 0 and 1 as $t$ does not go below $t_s$. Also note that the virtuality dependence of $\hat{q}$ is the same regardless of the mass of the quark. Other transport coefficients, namely the longitudinal drag $\hat{e}$, and the longitudinal diffusion $\hat{e}_2$, though present in MATTER, are not explored here, as their virtuality dependence is currently unknown. 

Given the importance of having a virtuality-dependent $\hat{q}(t)$ for light flavor observables, the main aim of this study is to investigate whether a virtuality-dependent $\hat{q}(t)$ can also affect heavy flavors. As no calculations of $\hat{q}$ simultaneously include its virtuality and heavy quark mass dependence --- i.e., there is no $\hat{q}(t,M)$ available in the literature --- the light-flavor $\hat{q}(t)$ is used herein for both light and heavy flavor to estimate of the magnitude of $\hat{q}$. At very large virtualities $t\gg M^2$, we expect the heavy quark mass to play less of a role, and the following approximation 
\begin{eqnarray}
\hat{q}(t,M)\overset{t \gg M^2}{\simeq}\hat{q}(t)
\label{eq:qhat_tM}
\end{eqnarray}
 is taken throughout this study. To obtain $\hat{q}(t,M)$ in the future, the SCET scheme \cite{Abir:2015hta} should be combined with the virtuality-dependent approach \cite{Kumar:2019uvu}.
\subsubsection{Kinematic limits of the Sudakov form factor integral}\label{sec:hf_in_matter}

In order to determine the virtuality $t$ of the parent particle, as well as the momentum fraction of the $z$ of its decay products, we use the kinematics of the reaction $Q\to Q+g$ to first determine the minimum/maximum momentum fraction allowed for this process. Up to linear order in $t_0/t$ the $z$ limits are:
\begin{eqnarray}
z_{\rm min} &=& \frac{t_0}{t}+\frac{M^2}{M^2+t}+\mathcal{O}\left(\left(\frac{t_0}{t}\right)^2\right)\nonumber\\
z_{\rm max} &=& 1-\frac{t_0}{t}+\mathcal{O}\left(\left(\frac{t_0}{t}\right)^2\right),
\end{eqnarray} 
where $t_0$ is the lowest scale below which MATTER evolution for light flavor partons is physically applicable and is taken to be $t_0=1$ GeV$^2$. Requiring further that $z_{\rm max} > z_{\rm min}$ implies that $t$ has a lower bond $t_{\rm min}$ which is $t_{\rm min}=t_0\left(1+\sqrt{1+2M^2/t_0}\right)$. With these limits at hand,  determination of the virtuality $t$ is done by sampling the Sudakov form factor, which gives a probability for {\it no} decay:
\begin{eqnarray}
\Delta_{Q\to Q+g}\left(t,t_{\rm min}\right)= \exp\left[-\int^t_{t_{\rm min}} dt' \int^{z_{\rm max}}_{z_{\rm min}} dz \left(\frac{dN^{\rm vac}}{dz dt'} + \frac{dN^{\rm med}}{dz dt'}\right)\right].
\label{eq:sud_Delta}
\end{eqnarray}
Once the virtuality of the parent parton is determined the momentum fraction $z$ can be determined by sampling $\int^{z_{\rm high}}_{z_{\rm low}} dz \left(\frac{dN^{\rm vac}}{dz dt'} + \frac{dN^{\rm med}}{dz dt'}\right)$ between $z_{\rm low}\geq z_{\rm min}$ and $z_{\rm high}\leq z_{\rm max}$. 

Heavy quarks can be produced in the medium via $g\to Q+\bar{Q}$, though such a production is kinematically suppressed compared to light quark production via gluon decay. Unlike the case of  $Q\to Q+g$, the medium modifications to the process of $g\to Q+\bar{Q}$ have not yet been derived using SCET. Thus, this production process is approximated as follows \cite{Majumder:2013re,Cao:2017qpx}: 
\begin{eqnarray}
\frac{dN^{\rm vac}}{dz dt} + \frac{dN^{\rm med}}{dz dt } &=& \frac{\alpha_s(t)}{2\pi} \frac{P_{Q\gets g}(z)}{t}\left\{ 1+ \int^{\tau^+_Q}_0 d\tau^+\frac{\hat{q}}{z(1-z)t} \left[2-2\cos\left( \frac{\tau^+}{\tau^+_Q} \right)\right]\right\},\nonumber\\
\label{eq:g_QQ}
\end{eqnarray}  
where $P_{Q\gets g}(z)=T_R\left[z^2+(1-z)^2\right]$, while heavy quark mass corrections are neglected. Using Eq. (\ref{eq:g_QQ}), the probability for the gluon {\it not} to split into $Q+\bar{Q}$ can be determined. However, there are other channels contributing to the decay of the gluon and hence the total probability for gluon {\it not} to split is the product of the probabilities of gluon not splitting into pairs of gluons as well as heavy and light flavor quarks. Though not explicitly presented here, those channels are all accounted for in MATTER.

The kinematics of the $g\to Q+\bar{Q}$ decay limit the available phase space of this process. Indeed, assuming $M^2/t\ll 1$ and $t_0/t\ll 1$:
\begin{eqnarray}
z_{\rm min} &=& \frac{t_0+M^2}{t}+\mathcal{O}\left(\left(\frac{t_0+M^2}{t}\right)^2\right)\nonumber\\
z_{\rm max} &=& 1-\frac{t_0+M^2}{t}+\mathcal{O}\left(\left(\frac{t_0+M^2}{t}\right)^2\right).
\end{eqnarray} 
Requiring again that $z_{\rm max} > z_{\rm min}$ as well as $t>t_{\rm min}$, implies a $t_{\rm min}=2(M^2+t_0)$. The determination of $t$ and $z$ proceeds in the same way as for $Q\to Q+g$. 

Once MATTER determines that a splitting has happened, additional contributions, stemming from further $2\to 2$ scatterings, are calculated using LBT scattering rates, whose principles are described in Sec.~\ref{sec:LBT}. Though these medium-induced $2\to 2$ scatterings are not energetic enough to significantly alter the $t$ and $z$ of the parent/daughter partons in the shower, they may involve enough energy/momentum exchange to promote medium partons to become part of the jet shower, thus leaving dynamical sinks affecting the hydrodynamical equations of motion. Partons leaving the hydrodynamical descriptions are called ``recoil'' partons, whose back-reaction onto the hydrodynamical fields is currently being studied within the JETSCAPE Collaboration, but not herein.

\subsection{The linearized Boltzmann transport formalism}
\label{sec:LBT}
The linearized Boltzmann transport (LBT) simulation assumes that a small virtuality (see e.g. \cite{Cao:2020wlm} and references therein) has been reached before further interaction between partons and the QGP occurs. In that limit, LBT neglects the virtuality of the parton, using on-shell energy and momentum while calculating parton interactions with the medium, and restores it once a parton exits the LBT evolution. The main focus of this study is to inspect how energy-momentum exchange with the QGP affects the charm quark evolution. The description of light-flavor parton interactions with the QGP is found in Ref. \cite{JETSCAPE:2022jer}. The LBT formalism relies on solving the Boltzmann equation taking into account $2\to 2$ and $2\to 3$ processes. Specifically, the evolution of the momentum and position distribution of a hard quark $Q$ with momentum $p_1$ is given by \cite{Bellac:2011kqa,Cao:2016gvr,Cao:2017hhk}:
\begin{eqnarray}
p^\mu_1 \partial_\mu f_1\left(x_1,p_1\right)&=&\mathcal{C}_{\rm el}[f_1]+\mathcal{C}_{\rm inel}[f_1]\\
\mathcal{C}_{\rm el}[f_1] &=& \frac{d_2}{2}\int dP_2 \int dP_3 \int dP_3 (2\pi)^4\delta^{(4)}\left(p_1+p_2-p_3-p_4\right)\left\vert \mathcal{M}_{1,2\to3,4}\right\vert^2 \lambda_2\left(s,t,u\right) \nonumber\\
&\times& \left\{f_3\left({\bf p}_3\right)f_4\left({\bf p}_4\right)\left[1\pm f_1\left({\bf p}_1\right)\right]\left[1\pm f_2\left({\bf p}_2\right)\right]-f_1\left({\bf p}_1\right)f_2\left({\bf p}_2\right)\left[1\pm f_3\left({\bf p}_3\right)\right]\left[1\pm f_4\left({\bf p}_4\right)\right]\right\}
\end{eqnarray}
where $C_{\rm el}$ is the $2\to 2$ collision rate of the leading order (LO) perturbative QCD (pQCD) $1+2\to 3+4$ process, $d_2$ is the spin-color degeneracy of the incoming parton ``2'', $\int dP_i \equiv \int \frac{d^3 p_i}{(2\pi)^3 2 p^0_i}$ with $i=2,3,4$; while $\lambda_2\left(s,t,u\right)=\theta\left(s-2m^2_D\right) \theta\left(s+t-m^2_D\right)\theta\left(-t-m^2_D\right)$. Finally, the procedure to calculate LO pQCD matrix element can be found in Chapter 17.4 of Ref.~\cite{Peskin:1995ev}, for instance, with matrix elements given in \cite{Combridge:1978kx}. 

The medium-induced gluon radiation responsible for describing $2\to 3$ processes in $\mathcal{C}_{\rm inel}[f_1]$ uses the same higher twist formulation as that employed in MATTER presented in Eq.~(\ref{eq:hq_ht}). Using the latter, the average number of gluons emitted from a hard heavy quark, between time $t$ and $t + \Delta t$, is \cite{Wang:2001ifa,Zhang:2003wk,Majumder:2009ge}:
\begin{eqnarray}
\bar{N}^{\rm med}(t\to t + \Delta t) &\approx&\Delta t\int dz dk_{\perp}^2 \frac{dN^{\rm med}}{dz dk_{\perp}^2 dt}\\
\frac{dN^{\rm med}}{dz dk_{\perp}^2 dt}&=& \frac{2\alpha_s P(z)}{\pi k_{\perp}^4}\hat{q}\left(\frac{k_{\perp}^2}{k_{\perp}^2+z^2M^2}\right)^4\sin^2\left(\frac{t-t_i}{2\tau_f}\right).
\end{eqnarray}
As different successive emissions are independent, a Poisson distribution probability is employed, whereby the probability of emitting $n$ gluons is
\begin{equation}
\mathcal{P}(n)=\frac{\left(\bar{N}^{\rm med}\right)^{n}}{n!}e^{-\bar{N}^{\rm med}},
\end{equation}
while the probability of a total inelastic process is $\mathcal{P}_{\rm inel.}=1-e^{-\bar{N}^{\rm med}}$. The procedure to determine whether (and how many) elastic vs inelastic scatterings inside the QGP have occurred is explored in detail in Ref. \cite{JETSCAPE:2022jer}. The only undetermined coefficient in LBT is the strong coupling $\alpha_s$, which can be fixed to $\alpha^{({\rm eff})}_s=0.3$ as mentioned before, or implemented as running coupling $\alpha_s(\mu^2)$. Both of cases are explored below. The LBT framework also generates recoil partons in the same way as in MATTER described above. 

\section{Results}
\label{sec:results}
This work focuses on studying the interplay between the higher twist and the Boltzmann transport energy loss mechanisms of charm quarks in the QGP, with an emphasis on higher twist contribution since it is included for the first time for open heavy flavor in a multi-stage calculation. Since the current JETSCAPE computational setup doesn't have multiple jets propagating through the same medium simultaneously, the calculation of the nuclear modification factor $R_{AA}$ simplifies to:
\begin{eqnarray}
R^D_{AA}= \frac{\frac{d \sigma^D_{AA}}{d p_T}}{\frac{d \sigma^D_{pp}}{d p_T}}=\frac{\sum_{\ell} \frac{d N^D_{AA,\ell}}{d p_T}\hat{\sigma}_{\ell}}{\sum_{\ell} \frac{d N^D_{pp,\ell}}{d p_T}\hat{\sigma}_{\ell}}
\label{eq:R_AA}
\end{eqnarray}
where $\frac{dN^D_{AA}}{dp_T}$ and $\frac{dN^D_{pp}}{dp_T}$ are the multiplicity of D-mesons originating from A-A and p-p collisions in the experimentally given $p_T$ bin, respectively. The spectrum $\frac{d N^D_{AA,\ell}}{d p_T}$ is calculated utilizing our multi-stage model with PYTHIA generating the original hard scattering, MATTER accounting for virtuality ordered vacuum and in-medium splitting (see Sec.~\ref{sec:MATTER}), and LBT providing the medium-induced shower modification at low virtuality and high energy (see Sec.~\ref{sec:LBT}). The cross-section for producing the hard scattering process of the given range $\ell$ in transverse momentum $\hat{p}_T$ is $\hat{\sigma}_{\ell}$ ($\hat{p}_T$ is the transverse momentum of the exchanged parton in the hard scattering sampled by PYTHIA). Many $\hat{\sigma}_{\ell}$ are sampled, spanning a large kinematic range of the collision. The connection between the PYTHIA shower and the energy loss models is chosen to be $0.6$ fm/$c$ but the dependence of $R_{AA}$ on this quantity is found to be weak \cite{JETSCAPE:2022jer}. Roughly 10 million events are generated for one simulation and are evenly distributed among 400 fluid dynamical events giving rise to about 25000 events per fluid dynamical event. Hadronization is handled by the Colorless string hadronization routine \cite{JETSCAPE:2019udz}. Note that bottom quark energy loss is not accounted for in this study.

\begin{figure}[!h]
\begin{center}
\begin{tabular}{cc}
\includegraphics[width=0.495\textwidth]{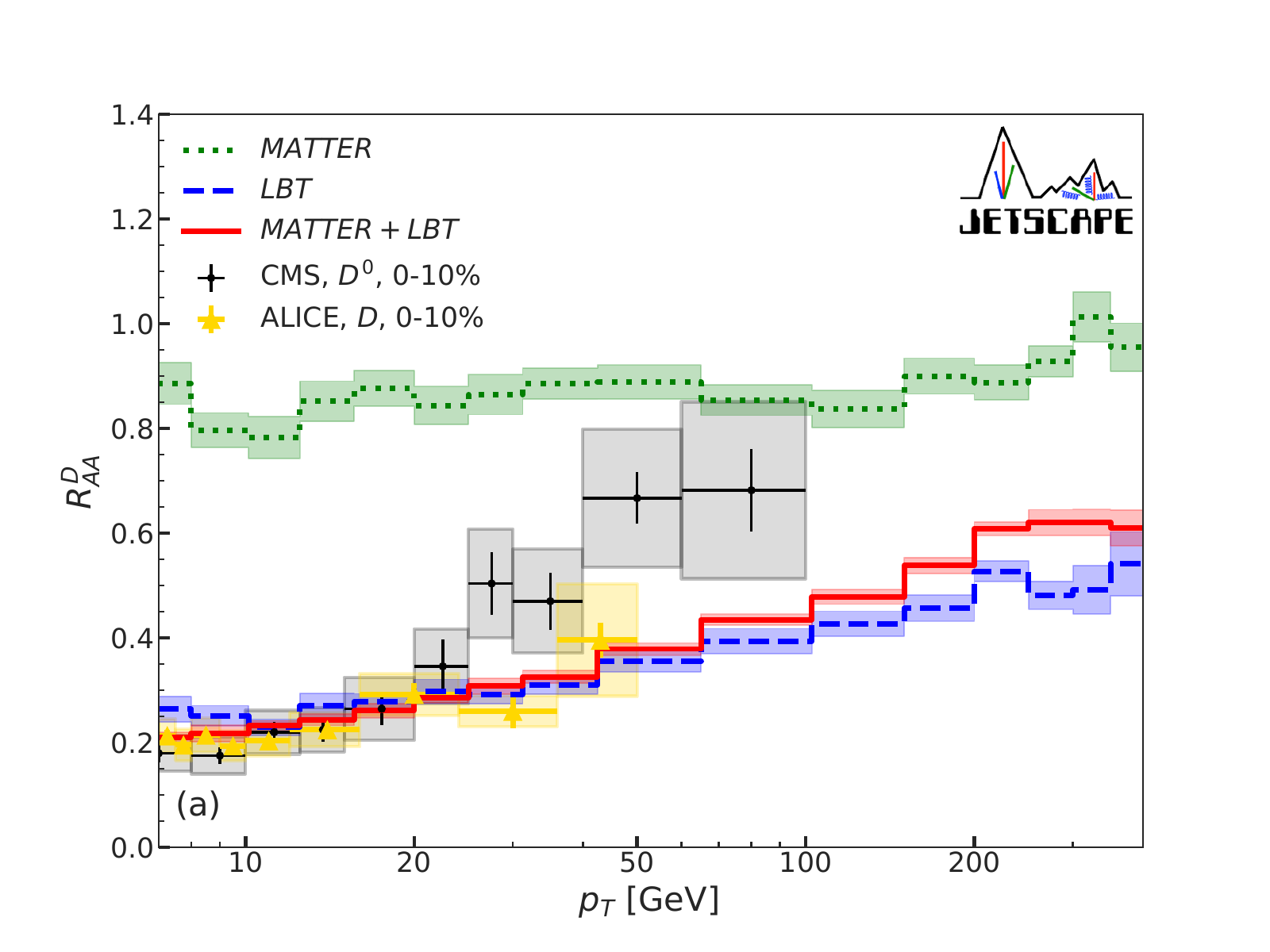} & \includegraphics[width=0.495\textwidth]{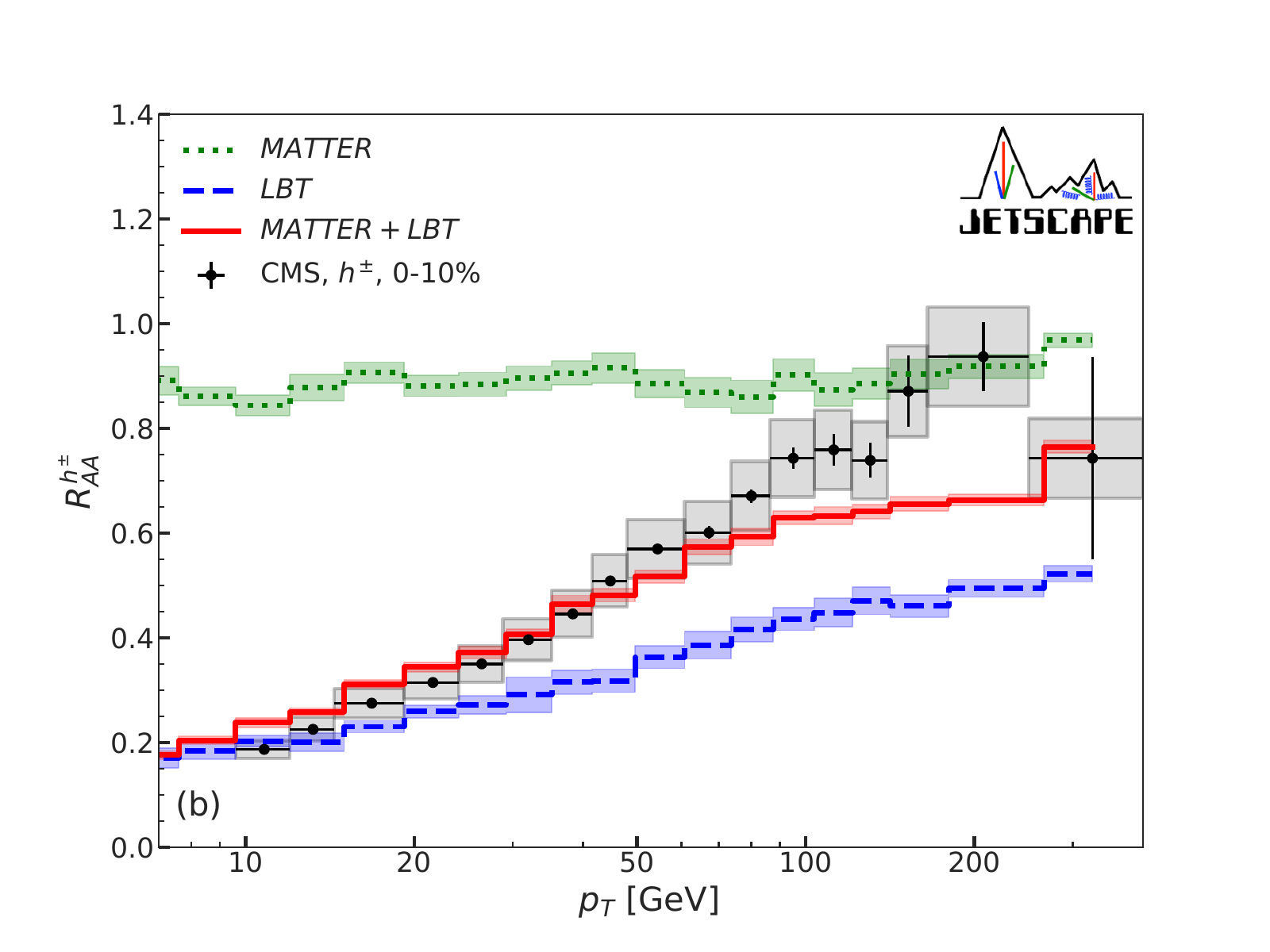}
\end{tabular}
\end{center}
\caption{(Color online) Nuclear modification factor for D-mesons (a) and charged hadrons (b) at the $\sqrt{s_{NN}}=5.02$ TeV Pb-Pb collisions at the LHC in the 0-10\% centrality. We set $c_1=10,\ c_2=100$ within the $\hat{q}(t)$ parametrization [see in Eq.~(\ref{eq:qhat_t})] for the MATTER alone and the MATTER+LBT curve. The other parameters for the MATTER+LBT curve is $t_s=4$GeV$^2$ found to best describe the $R_{AA}$ data. The p-p baseline for the LBT curve is calculated using PYTHIA whereas the p-p baseline for the MATTER and MATTER+LBT cases are calculated using MATTER vacuum \cite{JETSCAPE:2019udz}. Data taken from Ref.~\cite{CMS:2017qjw,CMS:2016xef,ALICE:2018lyv}.}
\label{fig:MAT_vs_LBT_all}
\end{figure}

Combining all the features of our calculation presented in Sec. \ref{sec:energy_loss}, namely a multi-stage simulation, a virtuality ($t$)-dependent $\hat{q}$, i.e. $\hat{q}(t)$, and a running $\alpha_s(\mu^2)$, results in the behavior seen in Fig. \ref{fig:MAT_vs_LBT_all}. On the left is the D-meson $R_{AA}$, while charged hadron $R_{AA}$ is on the right. Given the experimental uncertainties, an in-medium LBT or MATTER calculation alone has difficulty describing simultaneously charged hadrons and D-meson $R_{AA}$ over a wide $p_T$ range. A multi-stage calculation significantly improves the agreement to data, due to multiple contributing factors. In the following sections, MATTER and LBT simulations will first be studied in isolation.
This allows a deeper understanding of how each physical simulations affects $R_{AA}$, thus leading to an appreciation of how the improvement in Fig. \ref{fig:MAT_vs_LBT_all} is achieved within the combined simulation. 
\subsection{$R_{AA}$ from LBT}
\label{sec:results_lbt}
To obtain $R_{AA}$ using LBT as the sole energy loss mechanism, an initial parton distribution needs to be provided. One way to obtain this distribution is by using the PYTHIA vacuum shower. The latter is also used to provide the proton-proton baseline needed to calculate $R_{AA}$. Combining PYTHIA and LBT, two simulations were performed: one $\alpha^{\rm (eff)}_s=0.3$ serves as reference $R_{AA}$ calculation, while the other, using $\alpha_s(\mu^2)$, studies the effects of a running $\alpha_s$ on $R_{AA}$. 

%
\begin{figure}[!h]
\begin{center}
\begin{tabular}{cc}
\includegraphics[width=0.495\textwidth]{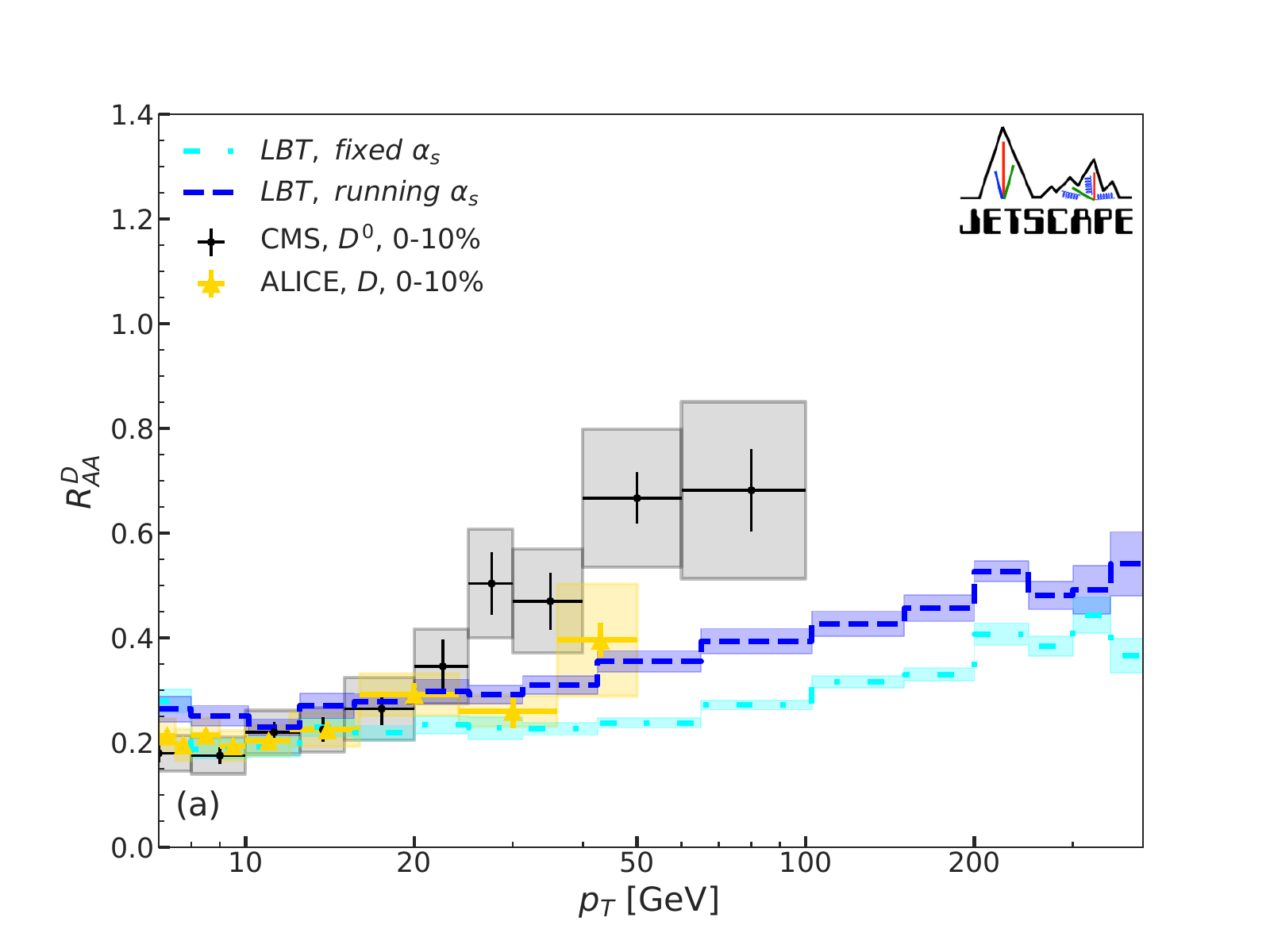} & \includegraphics[width=0.495\textwidth]{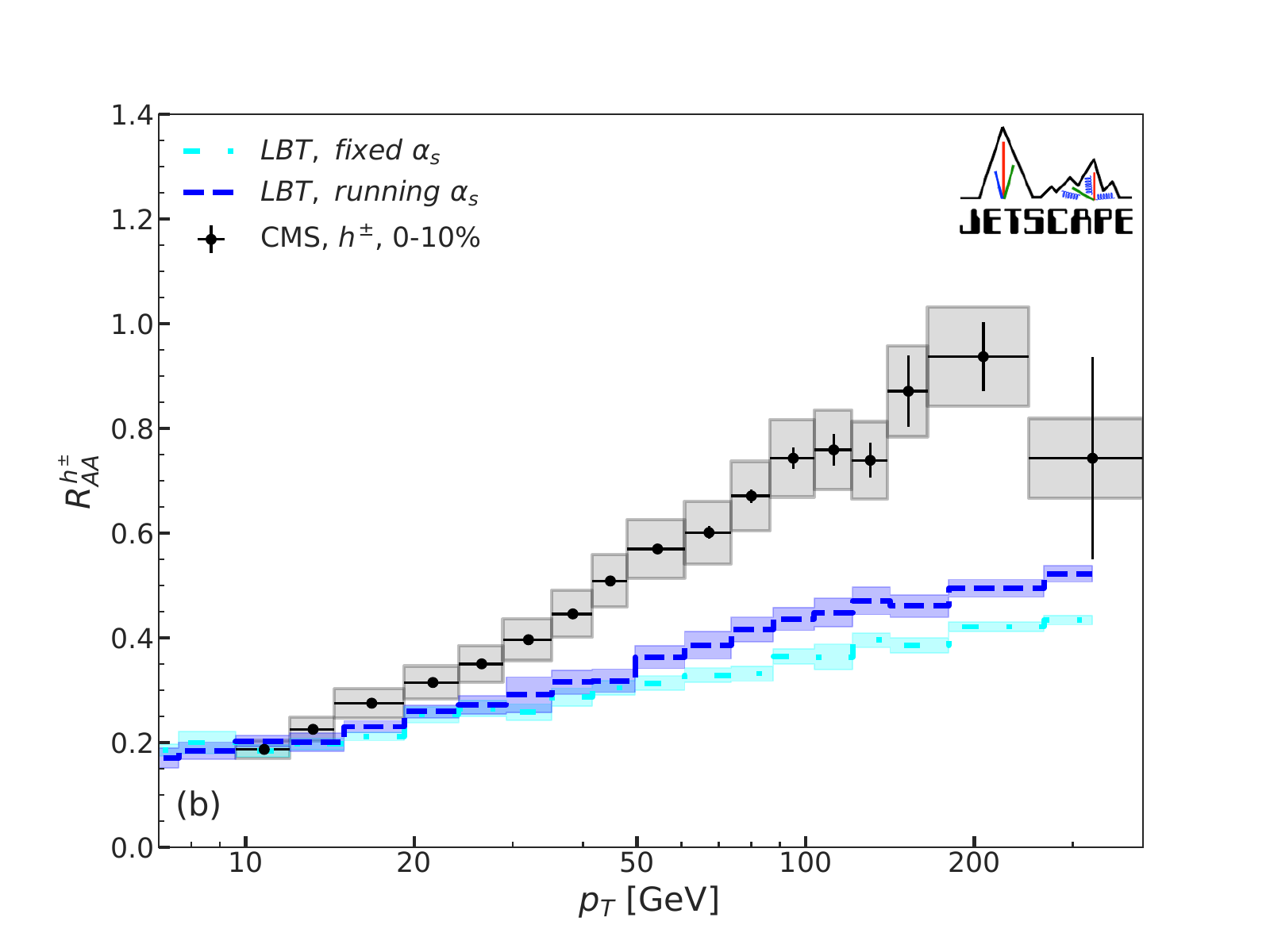}
\end{tabular}
\end{center}
\caption{(Color online) Nuclear modification factor for D-mesons (a) and charged hadrons (b) in $\sqrt{s_{NN}}=5.02$ TeV Pb-Pb collisions at the LHC at 0-10\% centrality. The p-p baseline is calculated using PYTHIA.}
\label{fig:LBT_comp}
\end{figure}
The results of these calculations are found in Fig. \ref{fig:LBT_comp}. Since these calculations rely on perturbation theory, solely results above $p_T\approx 7$ GeV/$c$ are shown. The calculations with constant $\alpha^{\rm (eff)}_s=0.3$ (dashed lines) generate significant energy loss at high $p_T$, producing an $R_{AA}$ slope that is inconsistent with data, for both charged hadrons and D-mesons. Including the effects of a running coupling $\alpha_s(\mu^2)$ (dotted lines) reduces the amount of parton interactions at high $p_T$, which improves the overall $R_{AA}$ slope to better mimic what is seen in experimental data. 

Note that past standalone LBT calculations can describe well experimental data on $R_{AA}$ and $v_2$ for both light and heavy quark hadrons, see Refs.~\cite{Cao:2017hhk, Xu:2018gux, Xing:2019xae, Cao:2020wlm} for example. These LBT calculations have a different treatment of initial spectra, bulk evolution, and hadronization procedures, compared the one shown here. Furthermore, different values of $\alpha_s$ for the interaction vertex connecting to thermal partons and jet partons were employed. However, Refs.~\cite{Cao:2017hhk, Xing:2019xae} have not considered how a medium-modified DGLAP showering mechanism (such as MATTER) can affect the subsequent Boltzmann transport evolution. The JETSCAPE framework is designed to connect different energy-loss schemes, such the medium-modified DGLAP evolution in MATTER and on-shell transport in LBT, thus going beyond studying them in isolation. Results from the standalone LBT as implemented in the JETSCAPE framework are shown here to illustrate the importance of a multi-stage in-medium jet shower evolution.

Except for $D^0$-meson $R_{AA}$ at high $p_T$, assuming that no energy loss occurs during the high virtuality showering of partons in a jet is an approximation that doesn't provide a good description of the data. Thus, the goal of the next section is to investigate how energy loss affects the high-virtuality portion of the shower simulated via the higher twist formalism in MATTER.

\subsection{$R_{AA}$ from MATTER}
\label{sec:results_matter}
%
\begin{figure}[!h]
\begin{center}
\begin{tabular}{cc}
\includegraphics[width=0.495\textwidth]{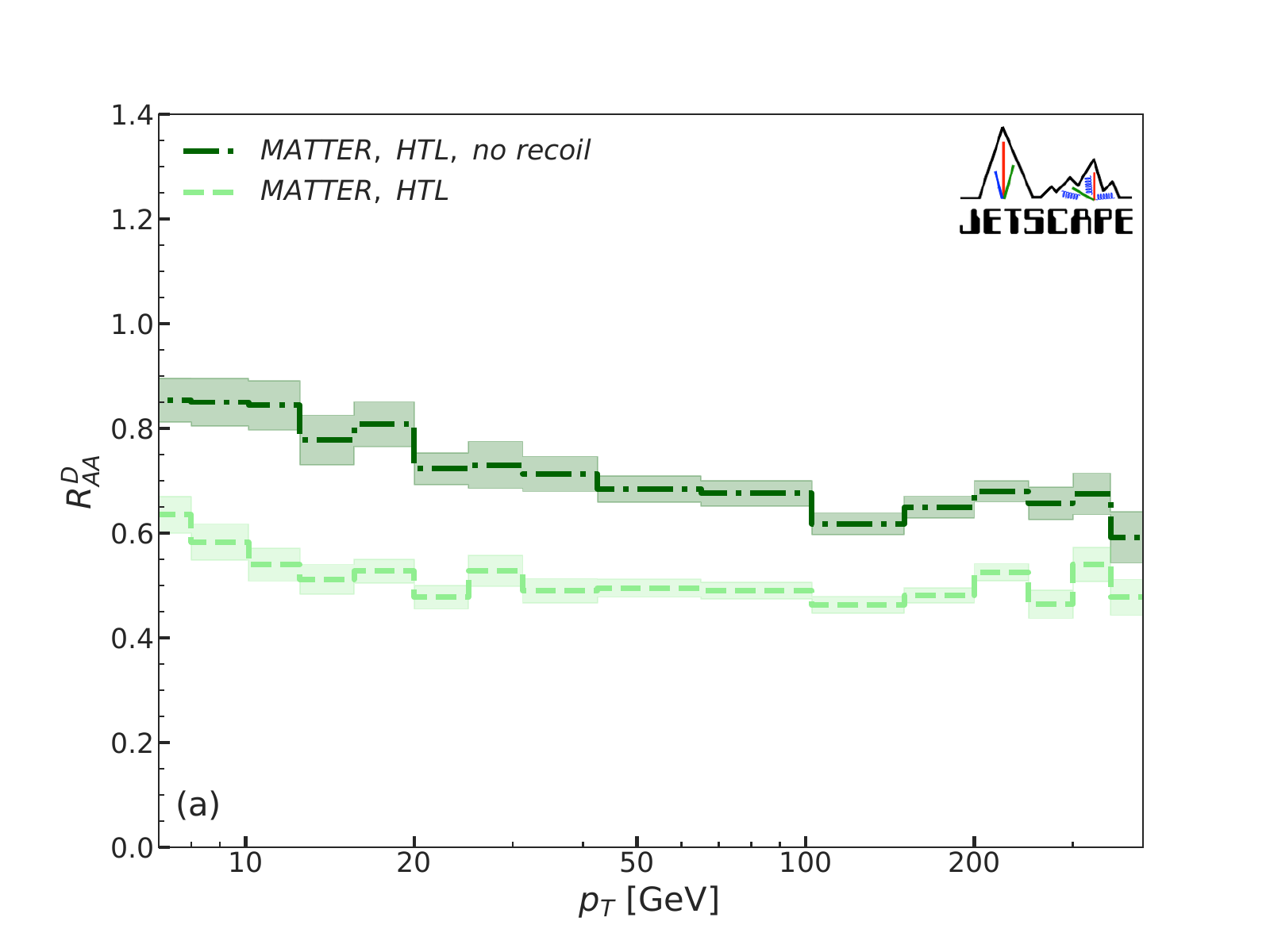} & \includegraphics[width=0.495\textwidth]{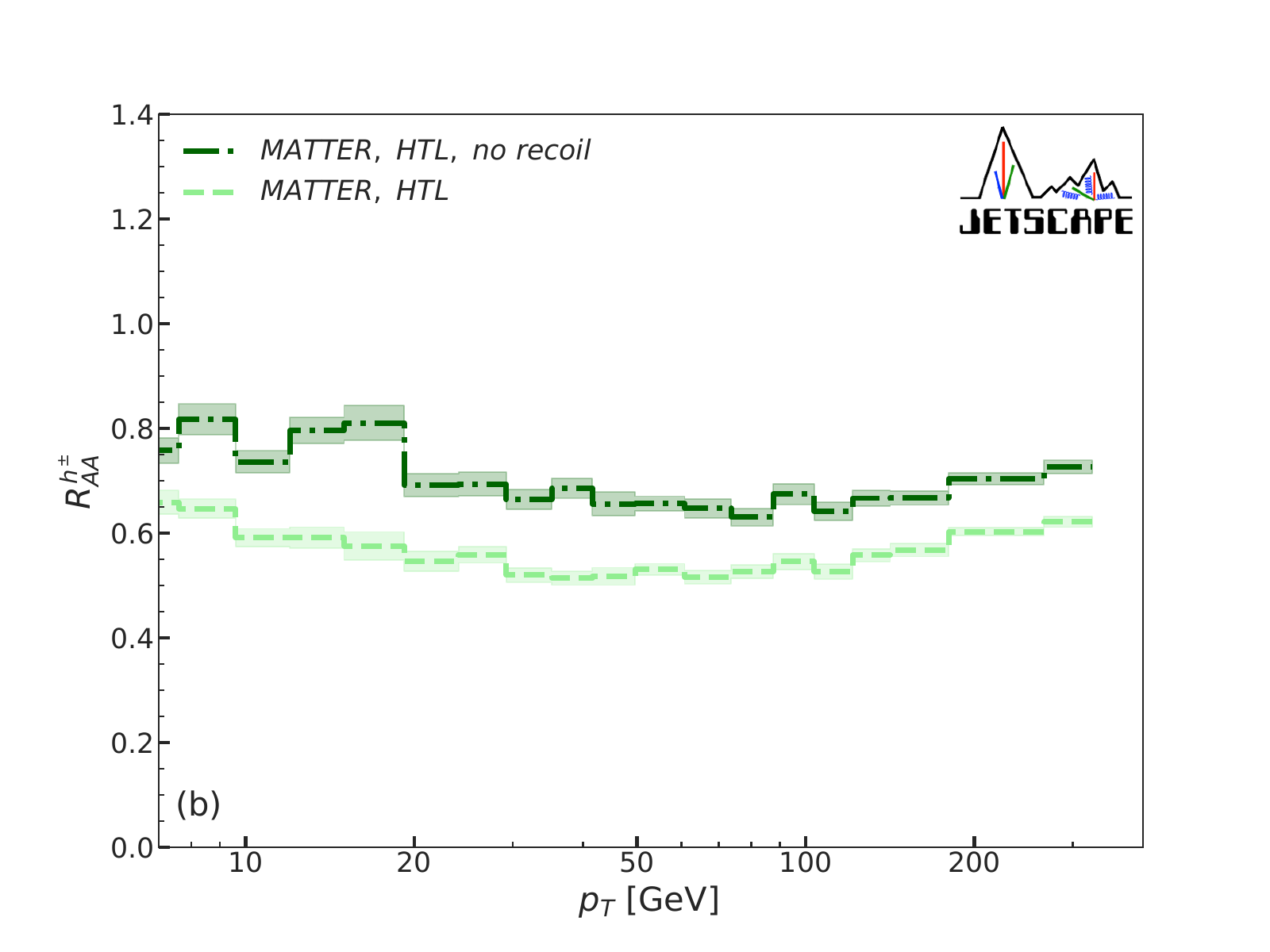}
\end{tabular}
\end{center}
\caption{(Color online) Nuclear modification factor for D-mesons (a) and charged hadrons (b) in $\sqrt{s_{NN}}=5.02$ TeV Pb-Pb collisions at the LHC at 0-10\% centrality. HTL denotes a Hard Thermal Loop calculation using $\hat{q}^{HTL}$, while no recoil refers to scattering processes being deactivated in MATTER.}
\label{fig:MATTER_comp_recoil}
\end{figure}
As MATTER is being used throughout the entire virtuality evolution herein, the higher twist formalism upon which it is based is employed until $t_s=1$ GeV$^2$. MATTER simulates the energy-momentum exchange between the partons of the medium and jet partons via two types of interactions. The first type of interaction is medium-induced inelastic radiation encapsulated in $\hat{q}$, a non-stochastic transport coefficient accounting for deviations from vacuum splittings. Elastic $2\to 2$ scatterings between jet and medium partons are treated stochastically. For each parton in the shower, the $2\to 2$ scattering rate is sampled. If a scattering occurs, the thermal parton involved can become part of the jet, leaving a negative contribution in the fluid, or become a source of energy-momentum to be deposited in the QGP. In Fig. \ref{fig:MATTER_comp_recoil}, the elastic and inelastic processes are studied in turn assuming a running $\alpha_s(\mu^2)$. 

Focusing on the result without $2\to2$ scatterings, labeled as no recoil in Fig.~\ref{fig:MATTER_comp_recoil}, one sees that including elastic scatterings leads to additional energy loss compared to that incurred via radiative processes alone. Unlike the LBT simulation where partons are long-lived and thus recoils are ever present, for a virtuality ordered shower in MATTER the importance of these elastic scatterings needs to be highlighted due to the highly variable lifetime of partons in the shower. Furthermore, the comparison between light and heavy flavor allows to appreciate how much these recoils affect partons of different masses. Our calculations show that heavy and light quarks are similarly affected, which can be an artifact of not using $\hat{q}(t,M)$, thus motivating its calculation by combining SCET of Ref.~\cite{Abir:2015hta} with Ref.~\cite{Kumar:2019uvu}. Outside of Fig.~\ref{fig:MATTER_comp_recoil}, $2\to 2$ scattering is always included in calculations containing MATTER.
\begin{figure}[!h]
\begin{center}
\begin{tabular}{cc}
\includegraphics[width=0.495\textwidth]{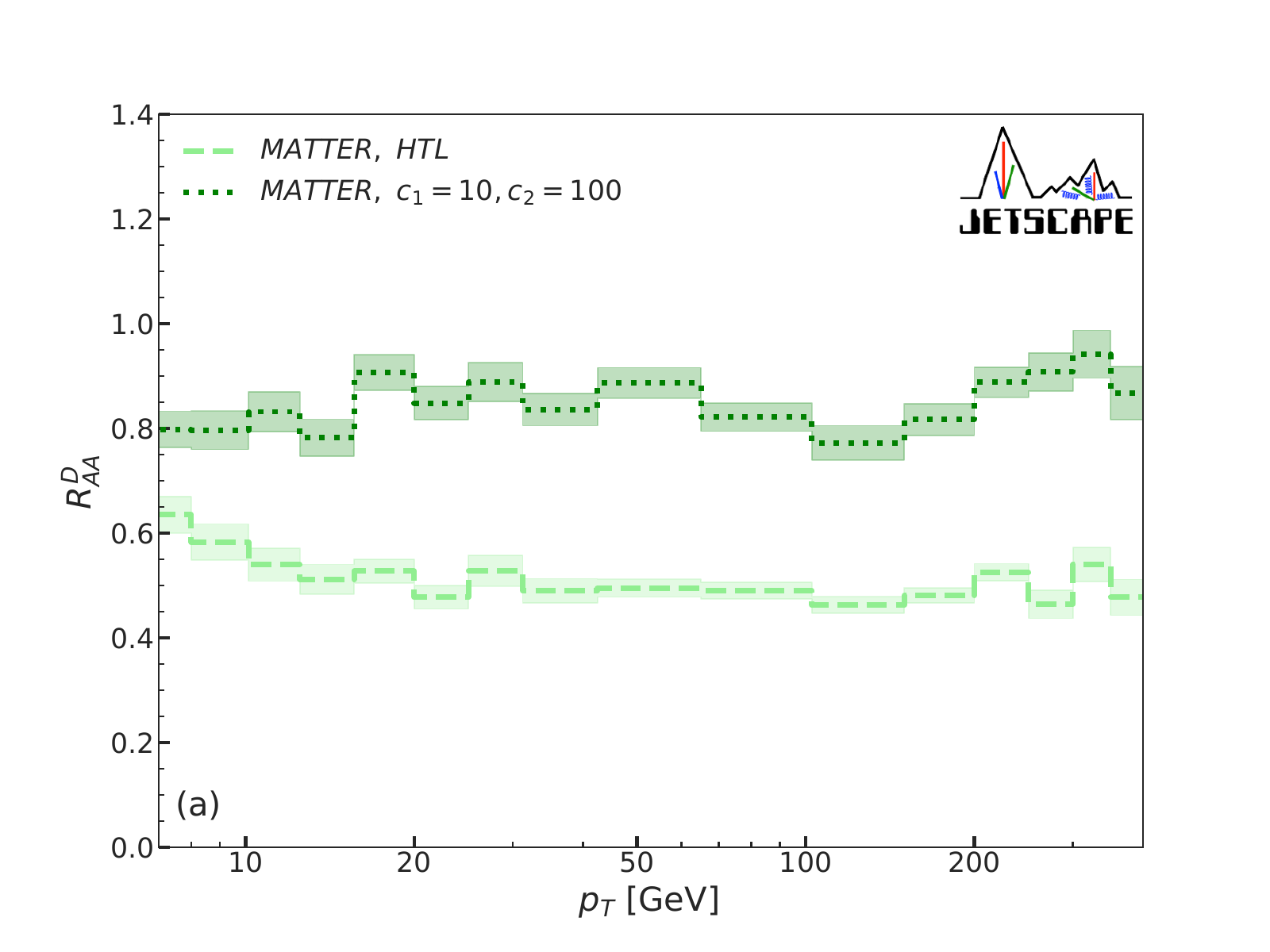} & \includegraphics[width=0.495\textwidth]{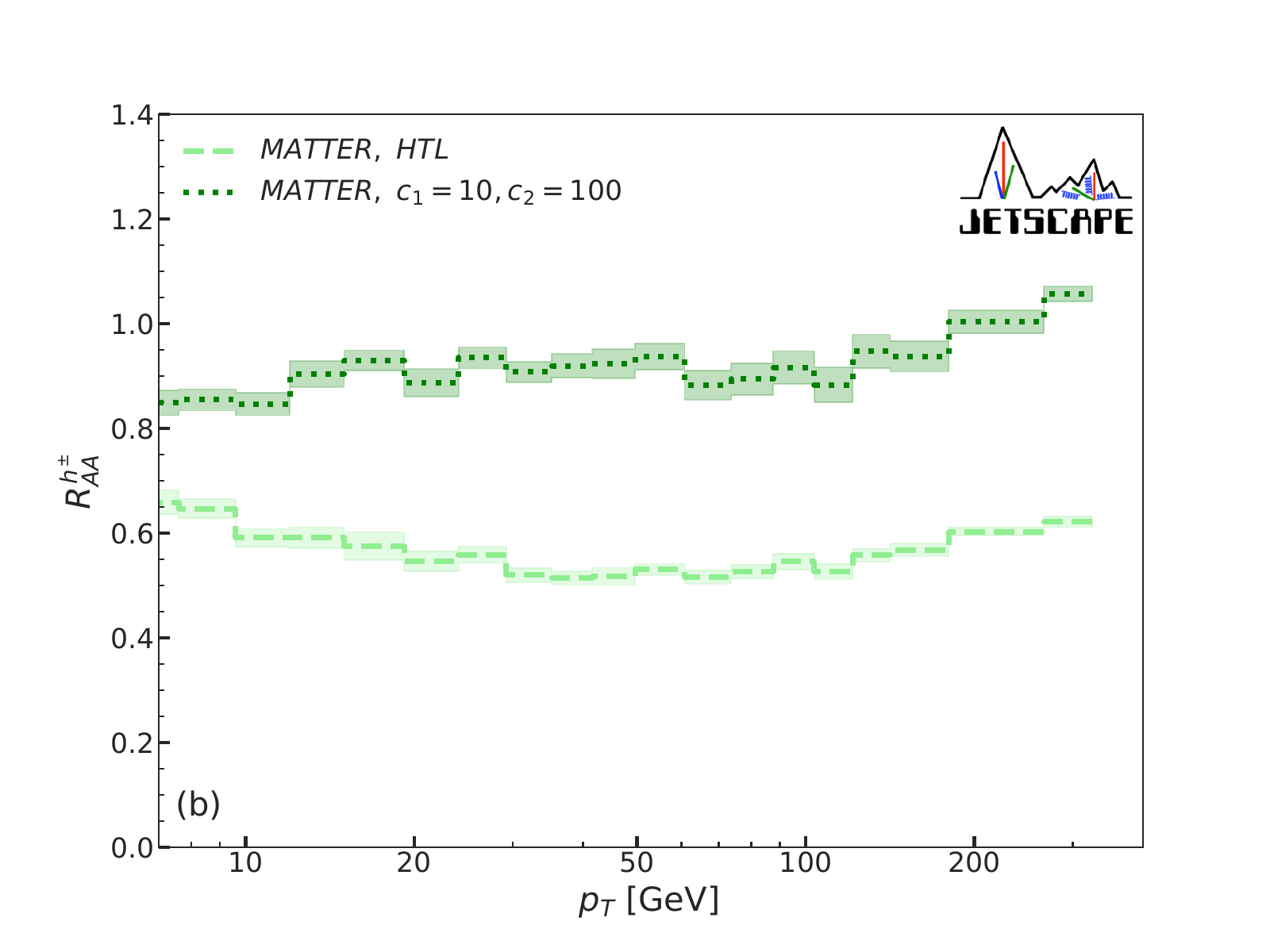}
\end{tabular}
\end{center}
\caption{(Color online) Nuclear modification factor for D-mesons (a) and charged hadrons (b) in $\sqrt{s_{NN}}=5.02$ TeV Pb-Pb collisions at the LHC at 0-10\% centrality. The difference between $\hat{q}^{HTL}$ and $\hat{q}(t)$ is significant especially at high $p_T$.}
\label{fig:MATTER_comp_qhat}
\end{figure}

As the virtuality dependent $\hat{q}(t)$ is smaller compared to the Hard Thermal Lool (HTL) result, the $R_{AA}$ tends to be much closer to $1$ for $\hat{q}(t)$ (dotted lines) compared to the one for $\hat{q}^{HTL}$ (dashed lines) as depicted in Fig. \ref{fig:MATTER_comp_qhat}. This effect is seen in both light and heavy flavor results at high $p_T$, as expected. 

The MATTER alone result is not to be compared with data, instead it gives a sense how different physics ingredients in MATTER affect its results, which are present in the overall comparison of MATTER+LBT $R_{AA}$ againts data.

\subsection{$R_{AA}$ from the combined MATTER and LBT simulation}
\label{sec:results_matter_lbt}
%
\begin{figure}[!h]
\begin{center}
\begin{tabular}{cc}
\includegraphics[width=0.495\textwidth]{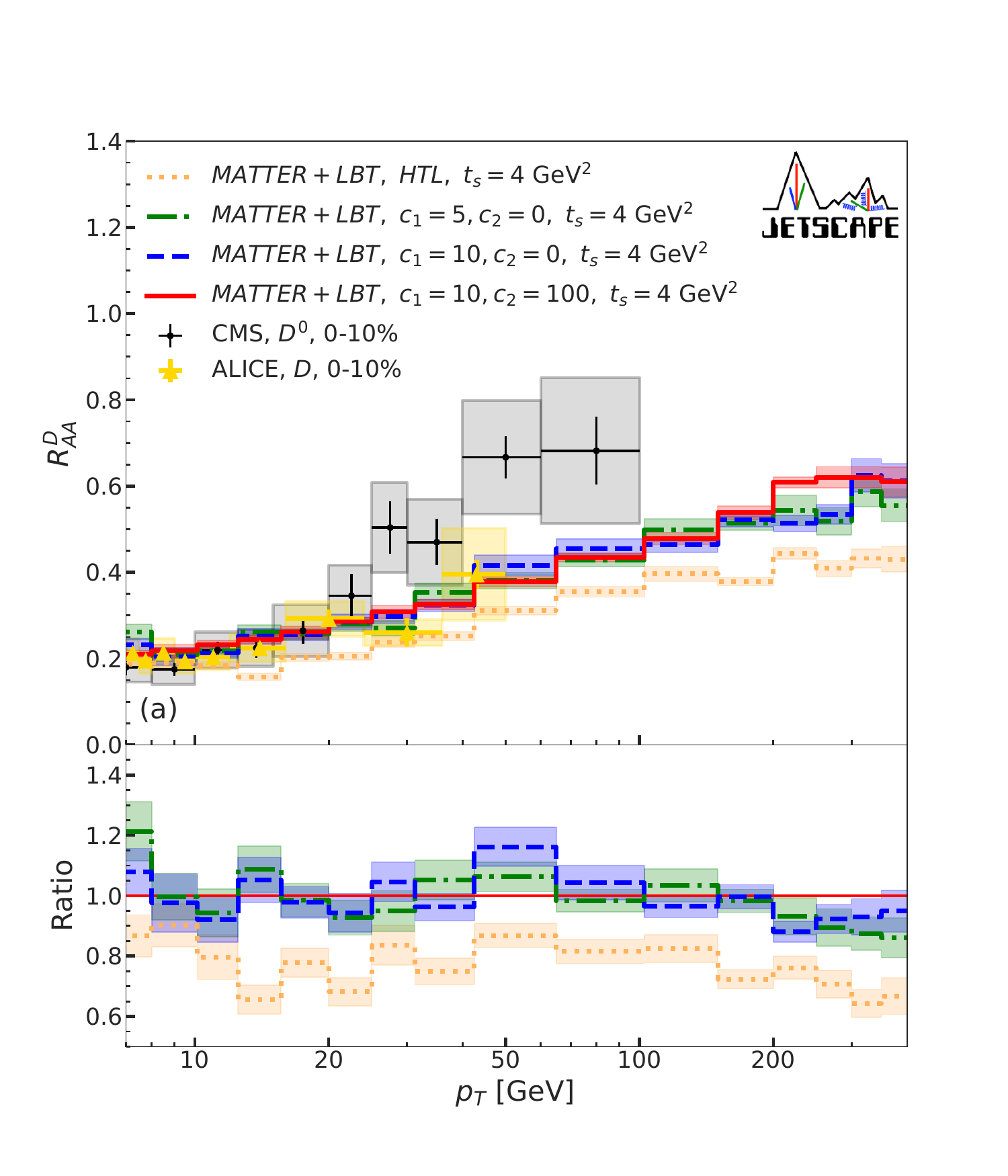} & \includegraphics[width=0.495\textwidth]{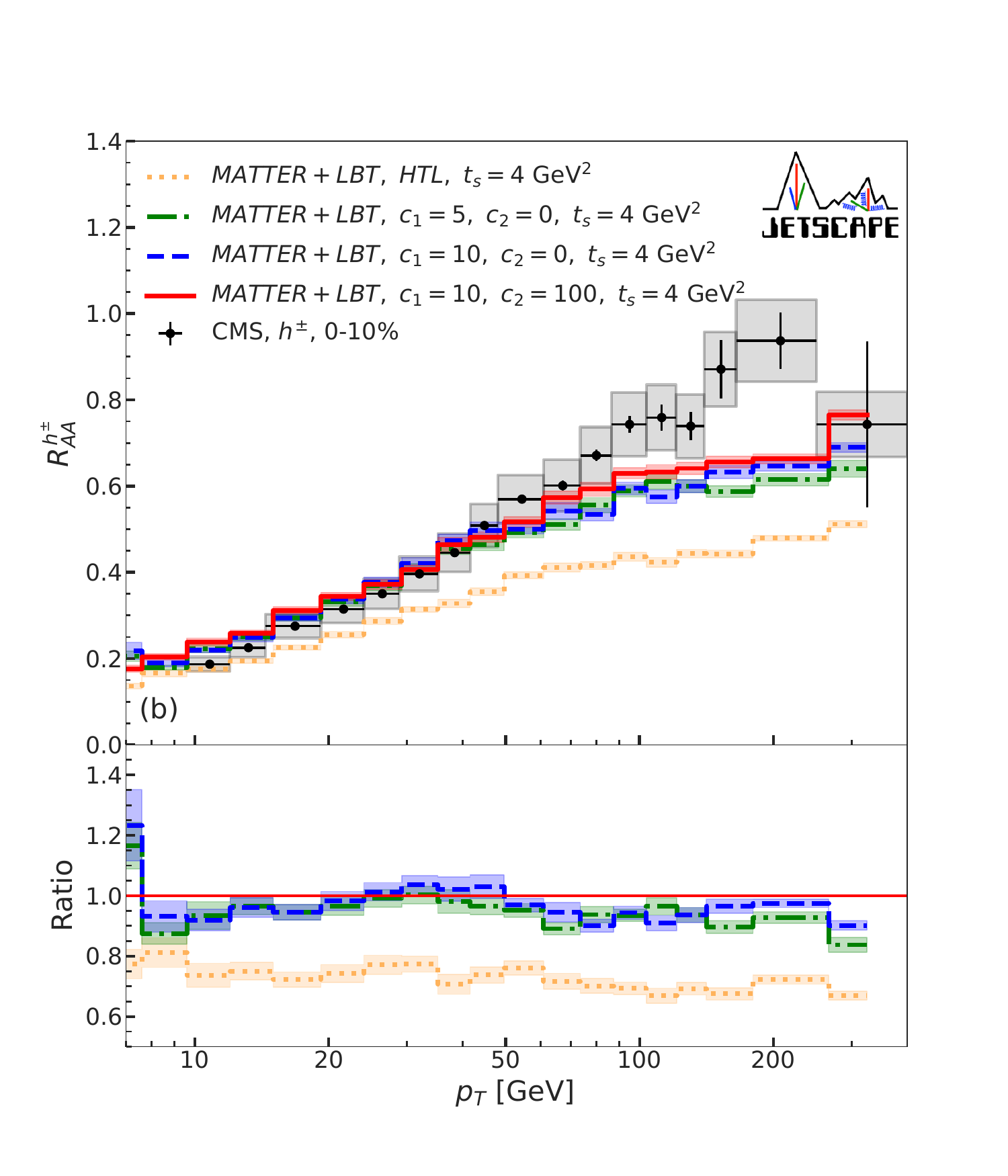}
\end{tabular}
\end{center}
\caption{(Color online) Nuclear modification factor for D-mesons (a) and charged hadrons (b) in $\sqrt{s_{NN}}=5.02$ TeV Pb-Pb collisions at the LHC at 0-10\% centrality. Here we are varying the parametrization of $\hat{q}(t)$ which is monotonically decreasing when $c_1$ and $c_2$ increase. The ratio in the bottom plots are taken with respect to the $c_1=10, c_2=100$ case with $\hat{q}(t)$ parametrization [see Eq.~(\ref{eq:qhat_t})]. A running $\alpha_s(\mu^2)$ is used in all calculations involving LBT.}
\label{fig:MATTER_LBT_comp_qhat}
\end{figure}
The combination of MATTER and LBT simulations is done by separating, in virtuality, the parton evolution in MATTER from that in LBT. The virtuality at which the switch is performed is a parameter, which for light flavor was tuned to $t_s=4$ GeV$^2$. 

A multi-stage $R_{AA}$ calculation using a virtuality-independent $\hat{q}^{HTL}$ alone shows an over suppression of $R_{AA}$ compared to data for both light and heavy flavors.  Additionally, the slope seen in the experimental data in the region $p_T\gtrsim 10$ GeV is steeper than what is obtained in our multi-stage $R_{AA}$ calculation using $\hat{q}^{HTL}$. A simple re-scaling of the overall normalization of $\hat{q}^{HTL}$ would not be enough to explain the slope seen in the data. In fact, a virtuality-dependent $\hat{q}$ whose value is suppressed as virtuality increases, such as that found Ref.~\cite{JETSCAPE:2022jer}, helps in this regard. Employing a virtuality-dependent $\hat{q}$ shows a significant effect on parton evolution not only in MATTER, but more importantly in the multi-stage MATTER+LBT evolution, affecting simultaneously light flavor and D-meson $R_{AA}$ seen in Fig.~\ref{fig:MATTER_LBT_comp_qhat}. It is the combination of a multi-stage simulation together with a virtuality-dependent $\hat{q}$ that is responsible for the agreement between the theoretical calculation and the data, in line with findings from the previous two sections. Fig.~\ref{fig:MATTER_LBT_comp_qhat} explores how different parameter values of $\hat{q}(t)$ in Eq.~(\ref{eq:qhat_t}) affect the $R_{AA}$, especially at high $p_T$. 

\subsubsection{Effects of $\hat{q}$ and $t_s$ on $R_{AA}$}
%
\begin{figure}[!h]
\begin{center}
\begin{tabular}{cc}
\includegraphics[width=0.495\textwidth]{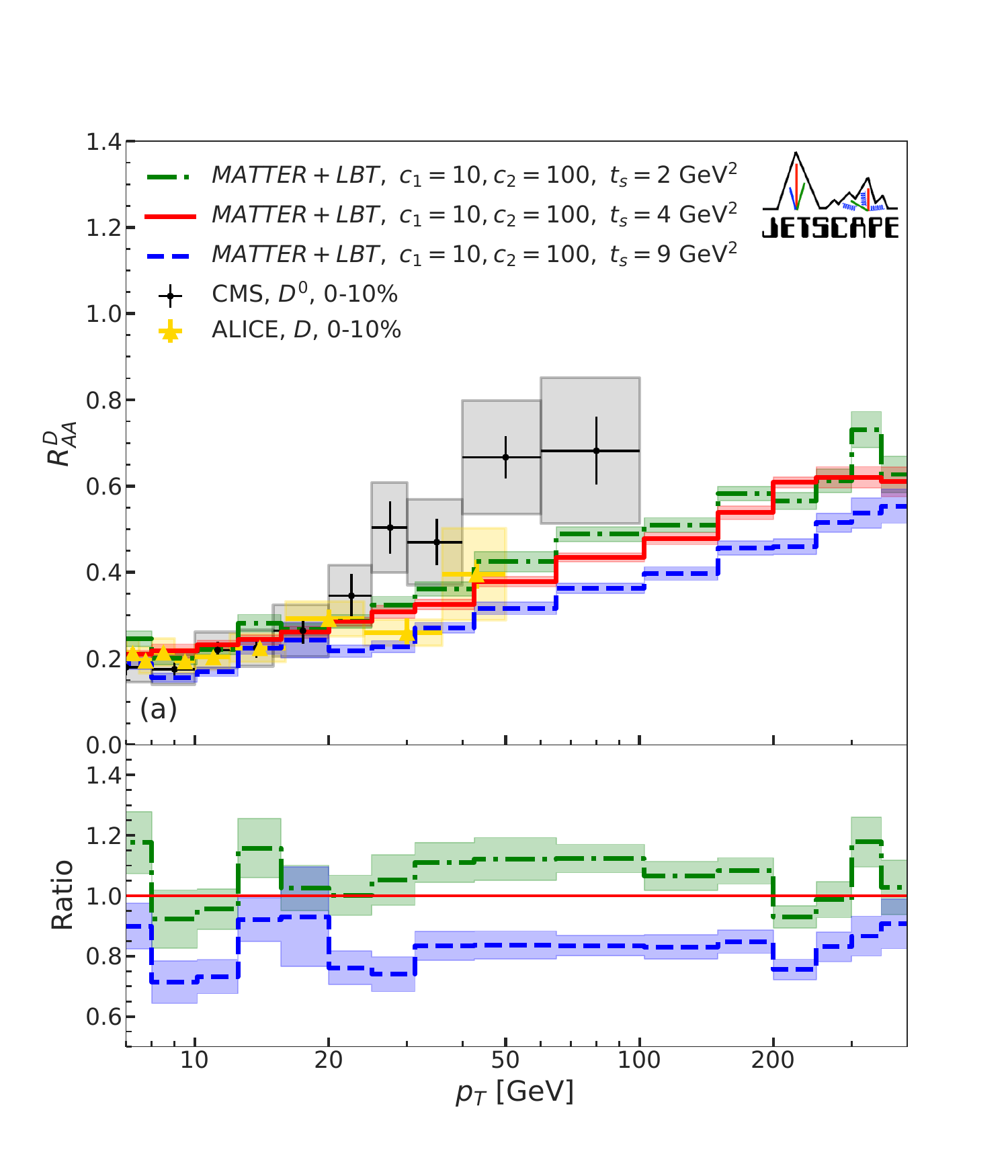} & \includegraphics[width=0.495\textwidth]{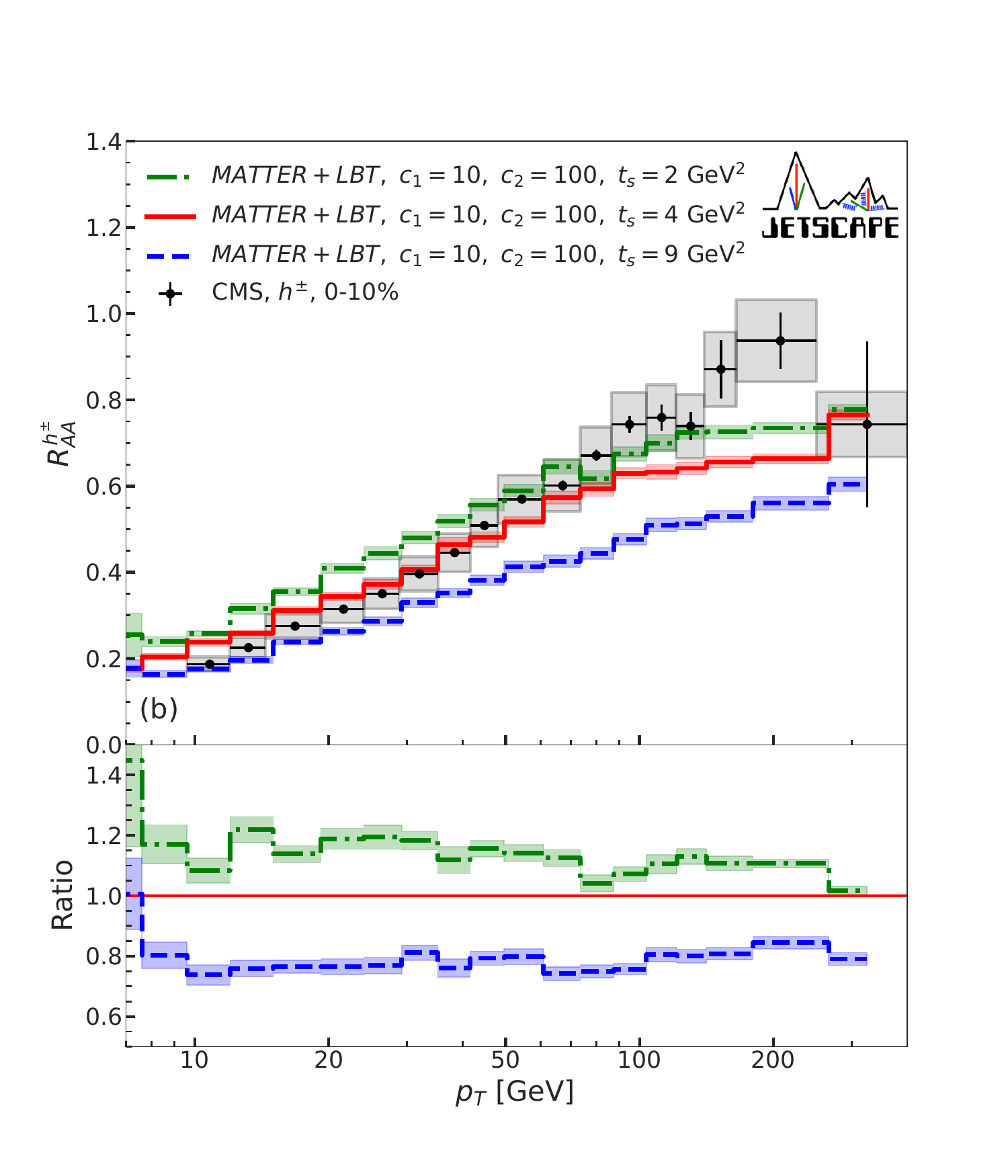}
\end{tabular}
\end{center}
\caption{(Color online) Nuclear modification factor for D-mesons (a) and charged hadrons (b) in $\sqrt{s_{NN}}=5.02$ TeV Pb-Pb collisions at the LHC at 0-10\% centrality. A bigger value of $t_s$ increases the effective length of LBT based energy-loss. The ratio in the bottom plots are taken with respect to the $c_1=10, c_2=100$ case of the $\hat{q}(t)$ parametrization. A running $\alpha_s(\mu^2)$ is used in all calculations involving LBT.}
\label{fig:MATTER_LBT_comp_Q0}
\end{figure}
Fig. \ref{fig:MATTER_LBT_comp_Q0} studies the effect of varying the switching scale $t_s$, with a larger $t_s$ implying a longer parton evolution in the LBT regime. The LBT mechanism generates significantly larger energy loss compared to the MATTER evolution, especially at low $p_T$, and thus the $t_s=9$~GeV$^2$ curve is close to a purely LBT simulation. Combining results from Figs.~\ref{fig:MATTER_LBT_comp_qhat} and \ref{fig:MATTER_LBT_comp_Q0}, we see that a parameter choice of $c_1=10, c_2=100, t_s=4$GeV$^2$ provides the best simultaneous description of the charged hadron and $D^0$ meson $R_{AA}$ data. To improve the description of $R_{AA}$ across all $p_T$, a Bayesian analysis of the $\hat{q}(t)$ parameter space is planned.
\subsubsection{Effects of gluon splitting to heavy quark pair on $R_{AA}$}
\begin{figure}[!h]
\begin{center}
\begin{tabular}{cc}
\includegraphics[width=0.495\textwidth]{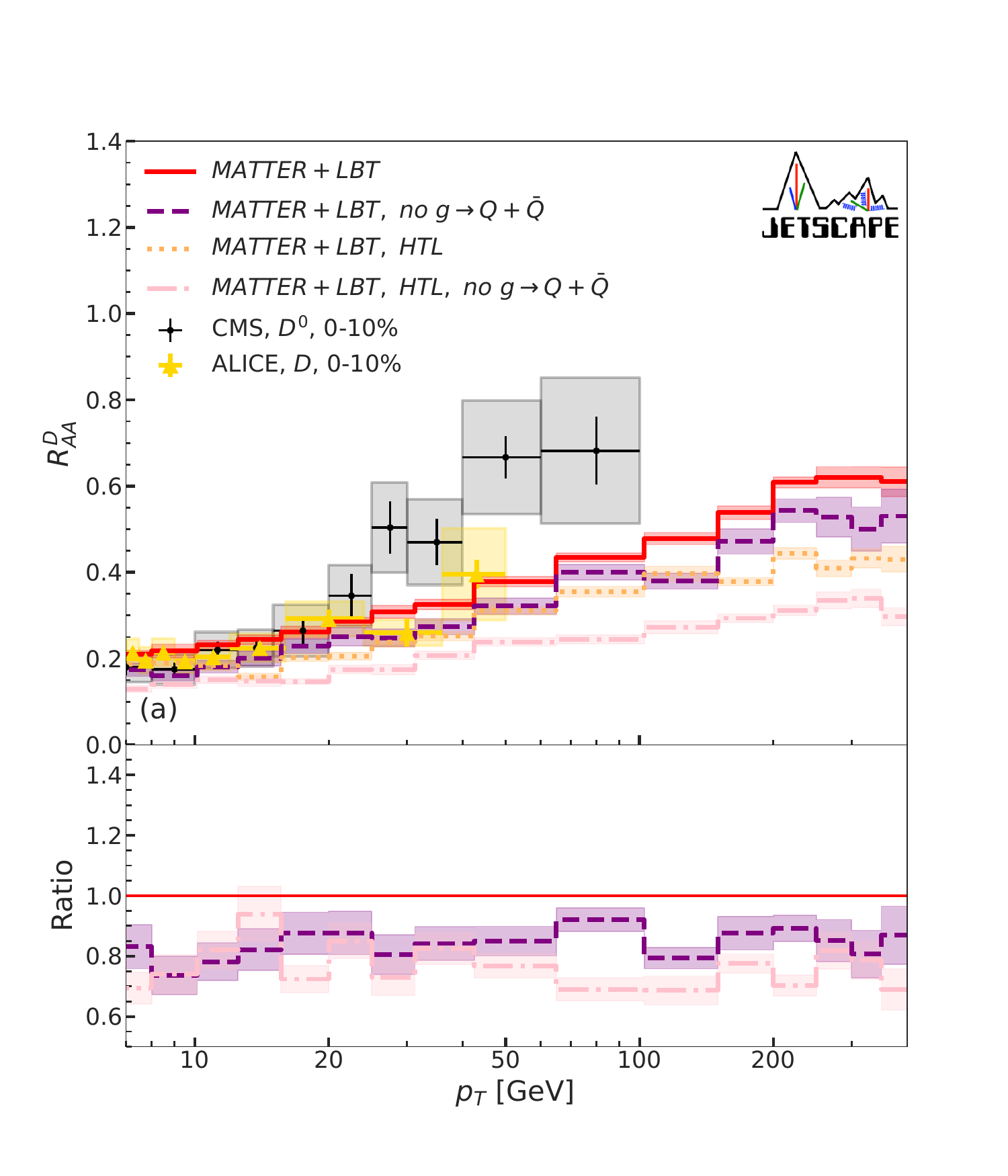} & \includegraphics[width=0.495\textwidth]{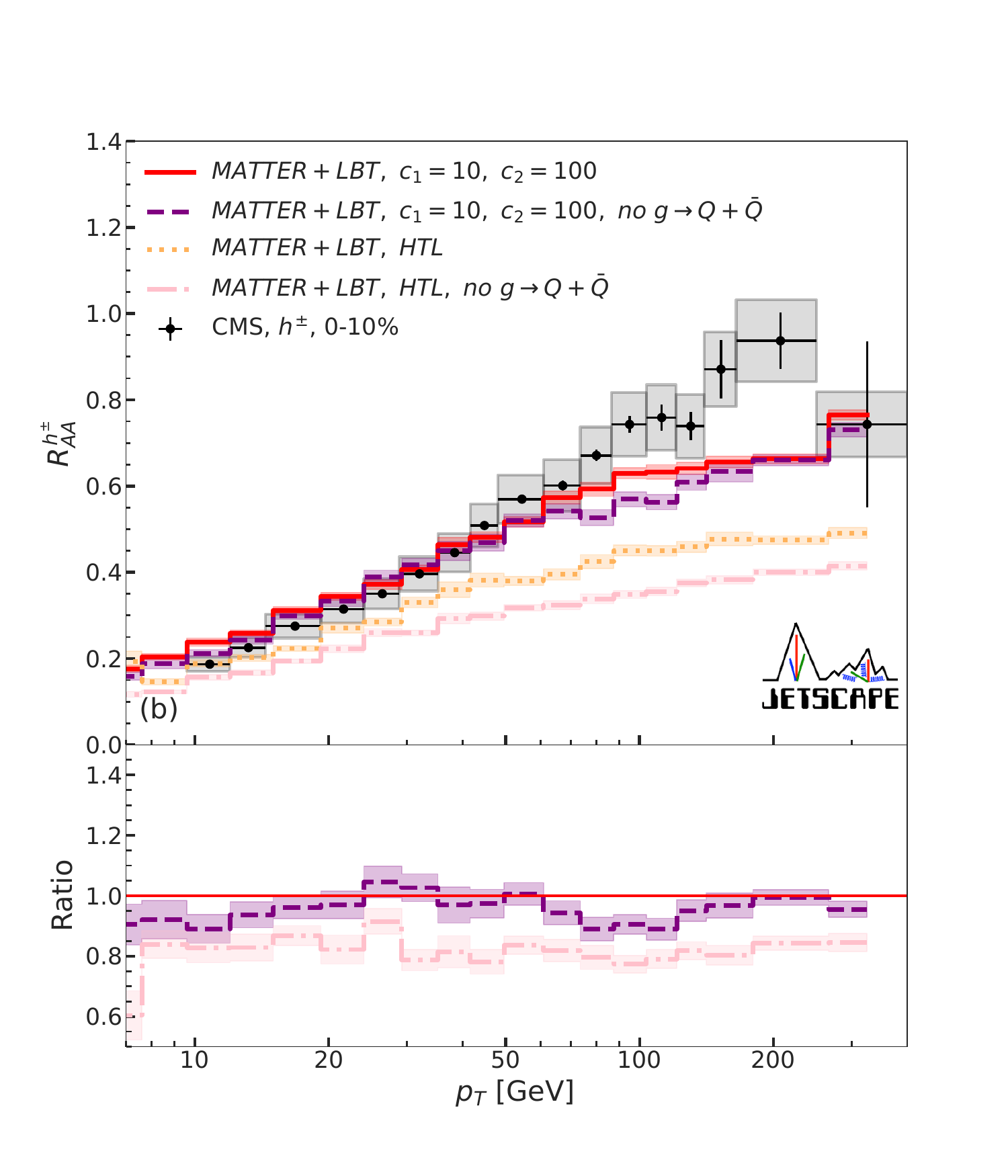}
\end{tabular}
\end{center}
\caption{(Color online) Nuclear modification factor for D-mesons (a) and charged hadrons (b) in $\sqrt{s_{NN}}=5.02$ TeV Pb-Pb collisions at the LHC at 0-10\% centrality. $c_1=10, \ c_2=100$ parameters values are employed in Eq.~(\ref{eq:qhat_t}). Ignoring the $g\rightarrow Q+\bar{Q}$ process in MATTER impacts the D meson $R_{AA}$, while it has a smaller effect on the charged hadron $R_{AA}$. A running $\alpha_s(\mu^2)$ is used in all calculations involving LBT. The dashed line in the ratio subplots divides MATTER+LBT no $g\to Q+\bar{Q}$ to MATTER+LBT, while the dotted-dashed line divides MATTER+LBT HTL no $g\to Q+\bar{Q}$ to MATTER+LBT HTL.}
\label{fig:MATTER_LBT_comp_gQQ}
\end{figure}
The novel physics ingredient that the present study allows to explore is the creation of heavy flavor through $g \rightarrow Q+\bar{Q}$ in MATTER, as presented in Sec. \ref{sec:hf_in_matter}. This process is best studied in a multi-scale simulation where the in-medium heavy quark creation and their subsequent evolution probes different virtuality regimes. Charmed quarks are the ideal candidate for this study, as their lighter mass (compared to bottom/top quarks) opens up the phase space for their in-medium dynamics, best highlighting the benefits of a multi-stage approach. To explore charm production from $g \rightarrow Q+\bar{Q}$, both the $D$ meson $R_{AA}$ and the charged hadron $R_{AA}$ are investigated using the combined MATTER and LBT simulation. As depicted in Fig.~\ref{fig:MATTER_LBT_comp_gQQ}, ignoring this process has a roughly $20\%$ impact on $D$ meson $R_{AA}$, while less than $10\%$ is seen for the charged hadron $R_{AA}$. Since we are only turning off the $g \rightarrow Q+\bar{Q}$ channel in MATTER, this has a smaller effect on the total charged hadron spectra, as $D$ meson contribution is subdominant. However, the contribution from gluon splitting to the total charm cross-section contribution is non-negligible, as a previous study using PYTHIA \cite{Norrbin:2000zc} also reports. The novel contribution the present simulation investigates is how different forms of $\hat{q}$ affect the $g\to Q+\bar{Q}$ heavy-flavor production. A larger $\hat{q}^{HTL}$ compared to $\hat{q}(t)$ reduces the parton virtuality, thus shrinking the phase space for $g\to Q+\bar{Q}$, which ultimately generates fewer charmed quarks and reduces the $R_{AA}$, for all curves but the solid one, as seen in Fig.~\ref{fig:MATTER_LBT_comp_gQQ} (a). Overall, Fig.~\ref{fig:MATTER_LBT_comp_gQQ} (a) provides the phenomenological importance to extend the SCET calculation of Ref.~\cite{Abir:2015hta} on the process $g\to Q+\bar{Q}$ in the future.
\subsubsection{$R_{AA}$ for $10-30\%$ and $30-50\%$ centrality}
%
\begin{figure}[!h]
\begin{center}
\begin{tabular}{cc}
\includegraphics[width=0.495\textwidth]{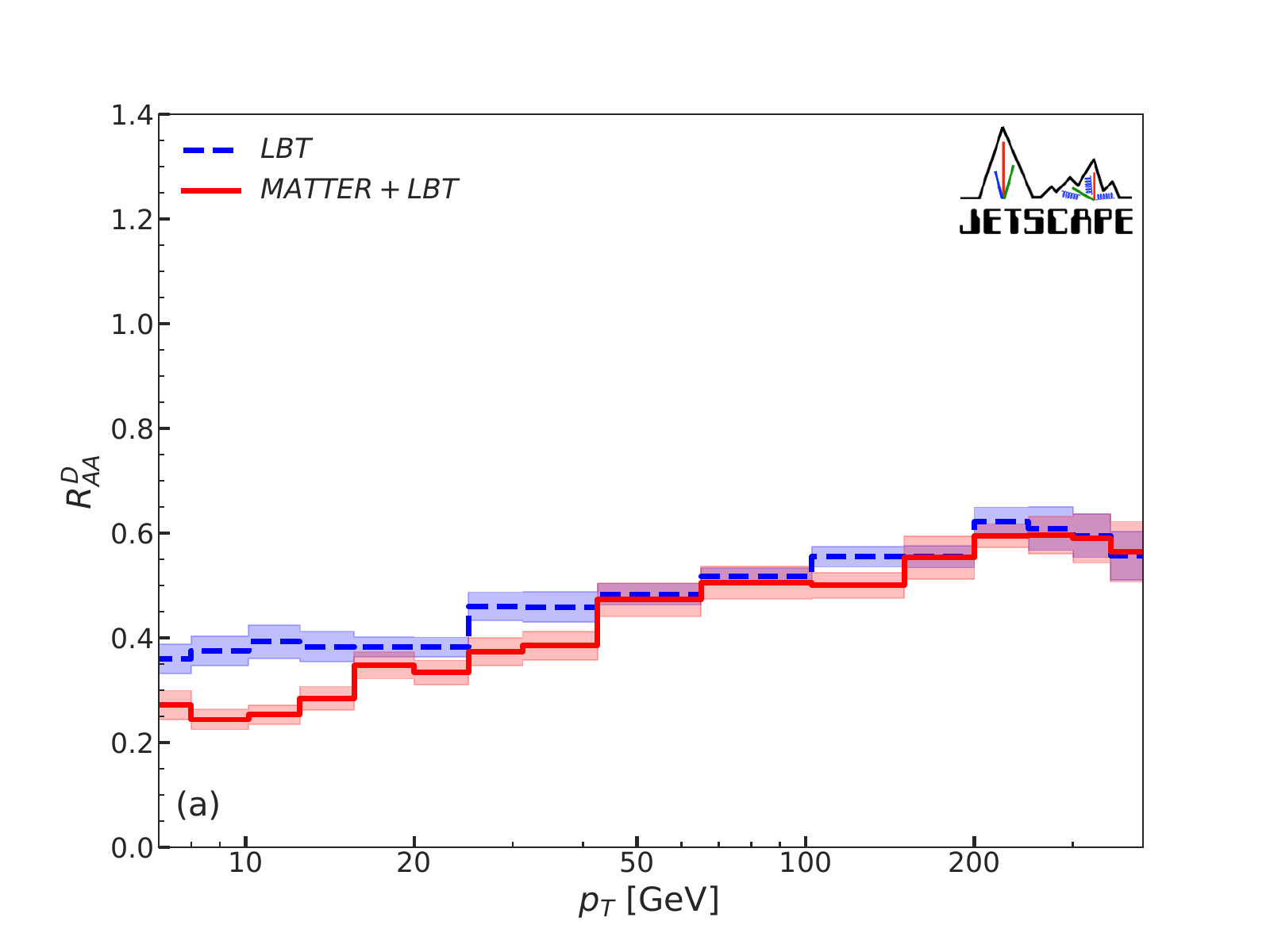} & \includegraphics[width=0.495\textwidth]{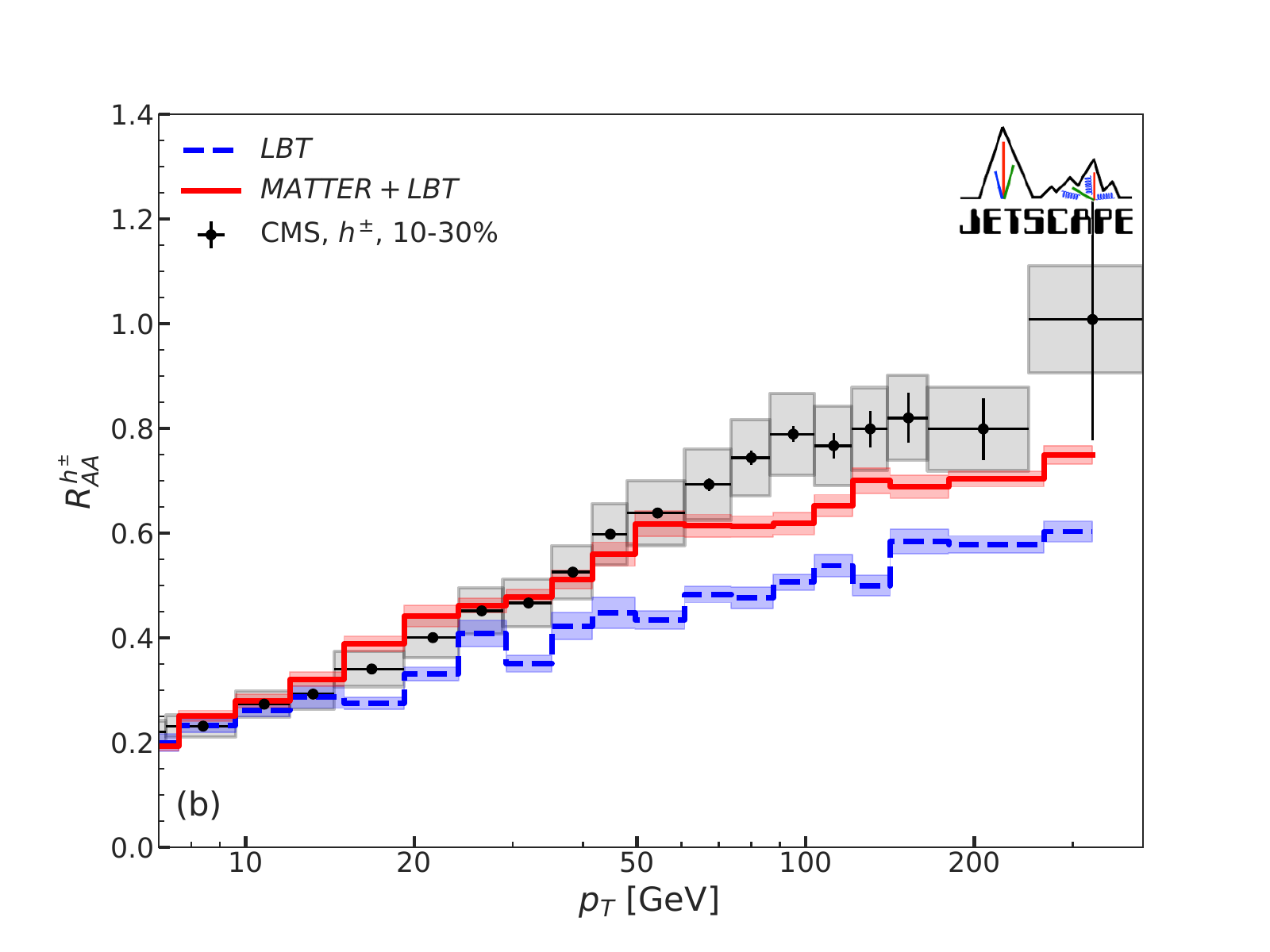}
\end{tabular}
\end{center}
\caption{(Color online) Nuclear modification factor for D-mesons (a) and charged hadrons (b) in $\sqrt{s_{NN}}=5.02$ TeV Pb-Pb collisions at the LHC at 10-30\% centrality. The parameters $c_1=10, \ c_2=100$ for the $\hat{q}(t)$ in Eq. (\ref{eq:qhat_t}) are chosen. A running $\alpha_s(\mu^2)$ is used in all calculations involving LBT.}
\label{fig:MATTER_LBT_10-30}
\end{figure}
\begin{figure}[!h]
\begin{center}
\begin{tabular}{cc}
\includegraphics[width=0.495\textwidth]{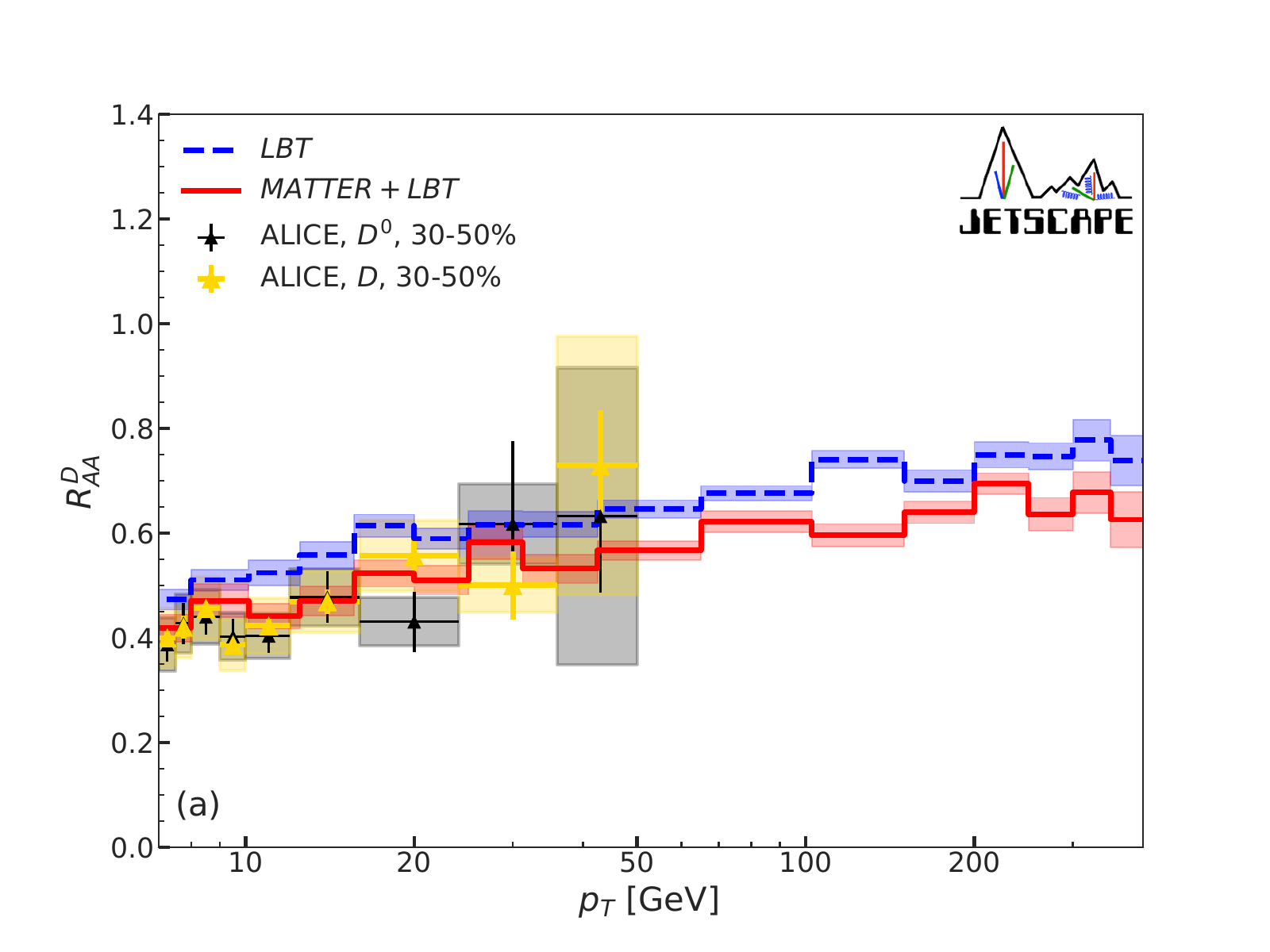} & \includegraphics[width=0.495\textwidth]{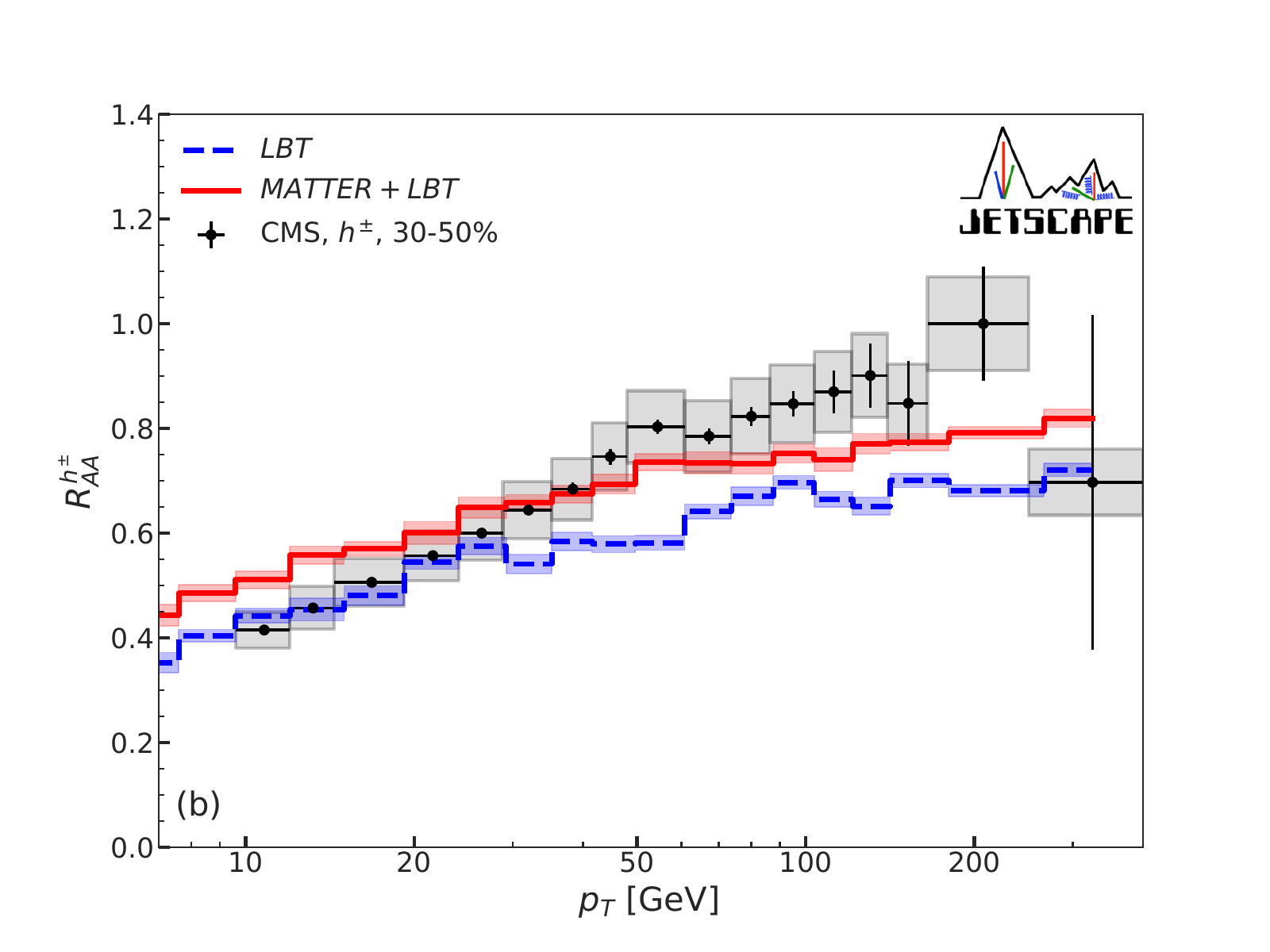}
\end{tabular}
\end{center}
\caption{(Color online) Nuclear modification factor for D-mesons (a) and charged hadrons (b) in $\sqrt{s_{NN}}=5.02$ TeV Pb-Pb collisions at the LHC at 30-50\% centrality. Here we choose $c_1=10, \ c_2=100$ for the $\hat{q}$ parametrization. A running $\alpha_s(\mu^2)$ is used in all calculations involving LBT.}
\label{fig:MATTER_LBT_30-50}
\end{figure}
\begin{figure}[!h]
\begin{center}
\begin{tabular}{cc}
\includegraphics[width=0.495\textwidth]{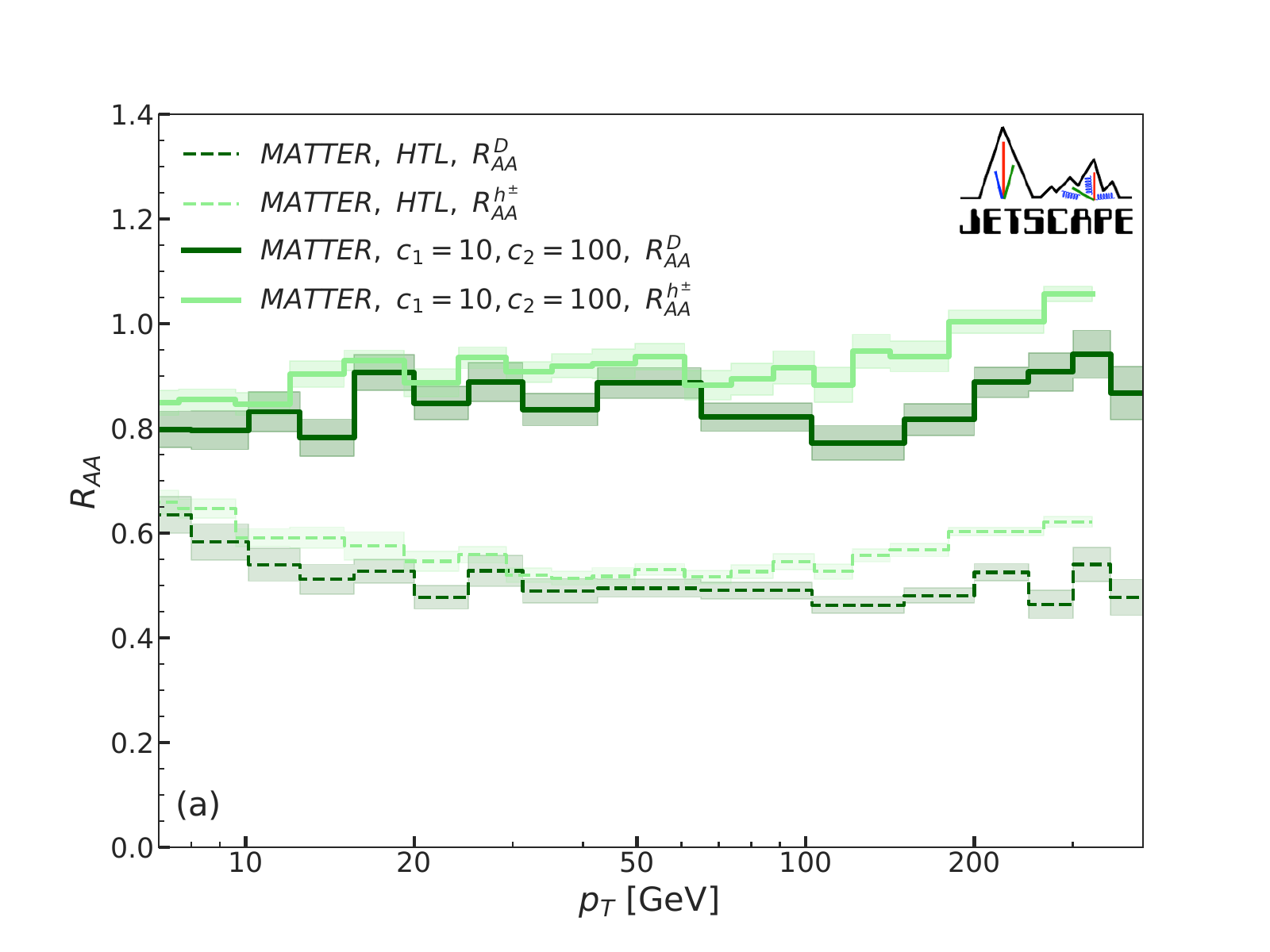} & \includegraphics[width=0.495\textwidth]{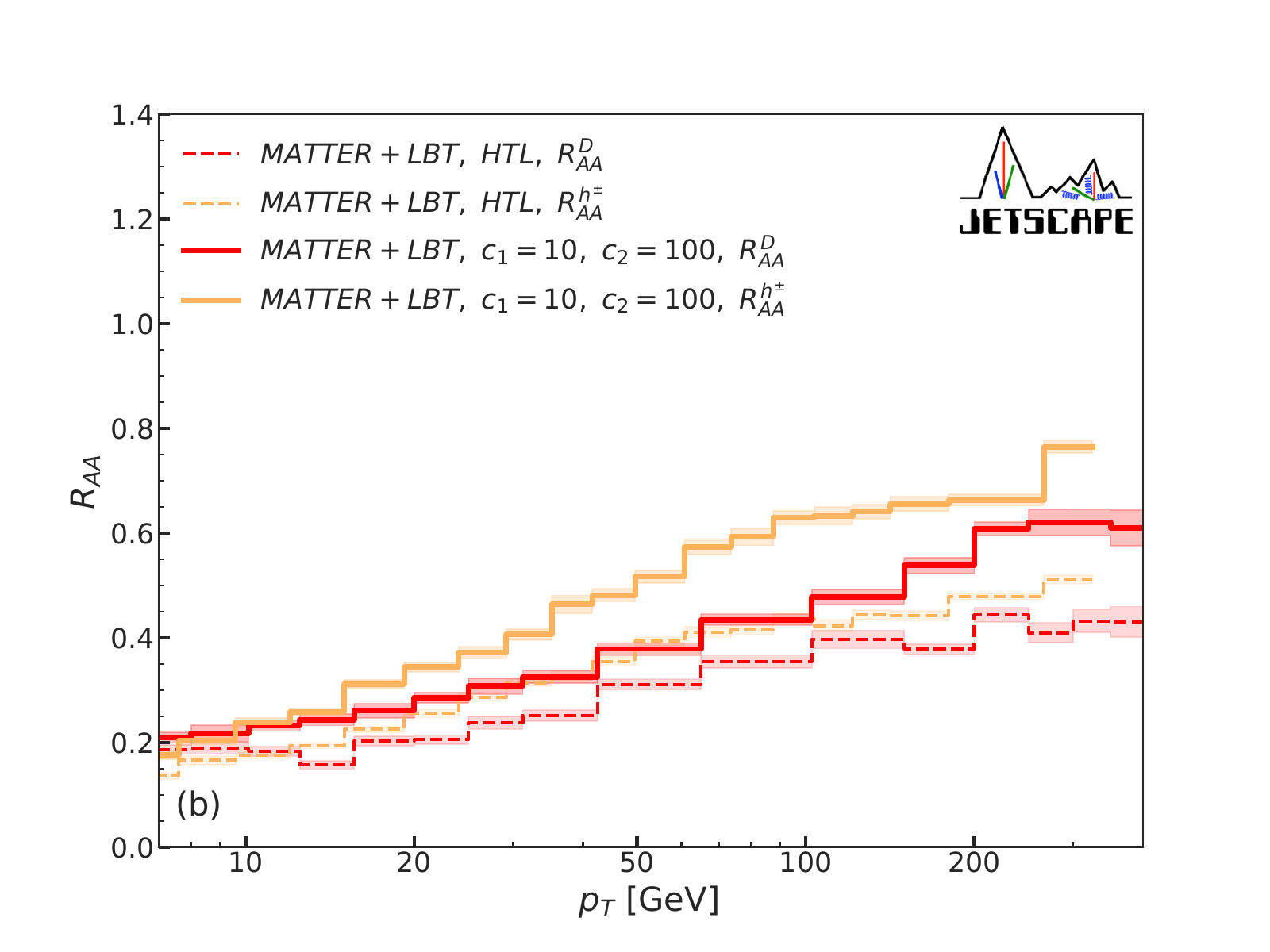}
\end{tabular}
\end{center}
\caption{(Color online) Nuclear modification factor for MATTER only simulations (a) and for MATTER+LBT simulations (b) in $\sqrt{s_{NN}}=5.02$ TeV Pb-Pb collisions at the LHC at 0-10\% centrality. $c_1=10, \ c_2=100$ parameters values are employed in Eq.~(\ref{eq:qhat_t}). A running $\alpha_s(\mu^2)$ is used in all calculations involving LBT.}
\label{fig:MATTER_MATTER+LBT_compare}
\end{figure}
\begin{figure}[!h]
\begin{center}
\begin{tabular}{cc}
\includegraphics[width=0.495\textwidth]{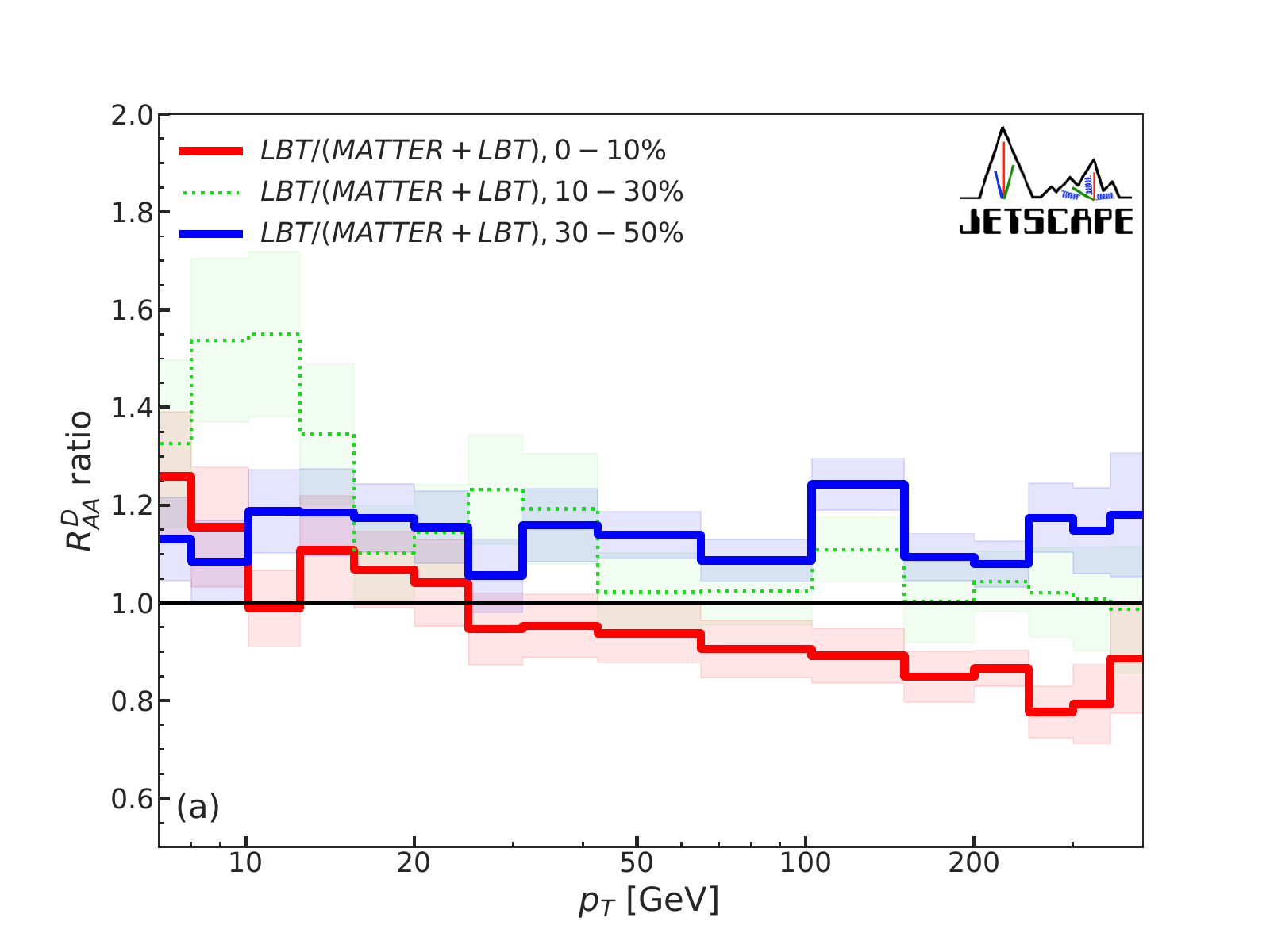} & \includegraphics[width=0.495\textwidth]{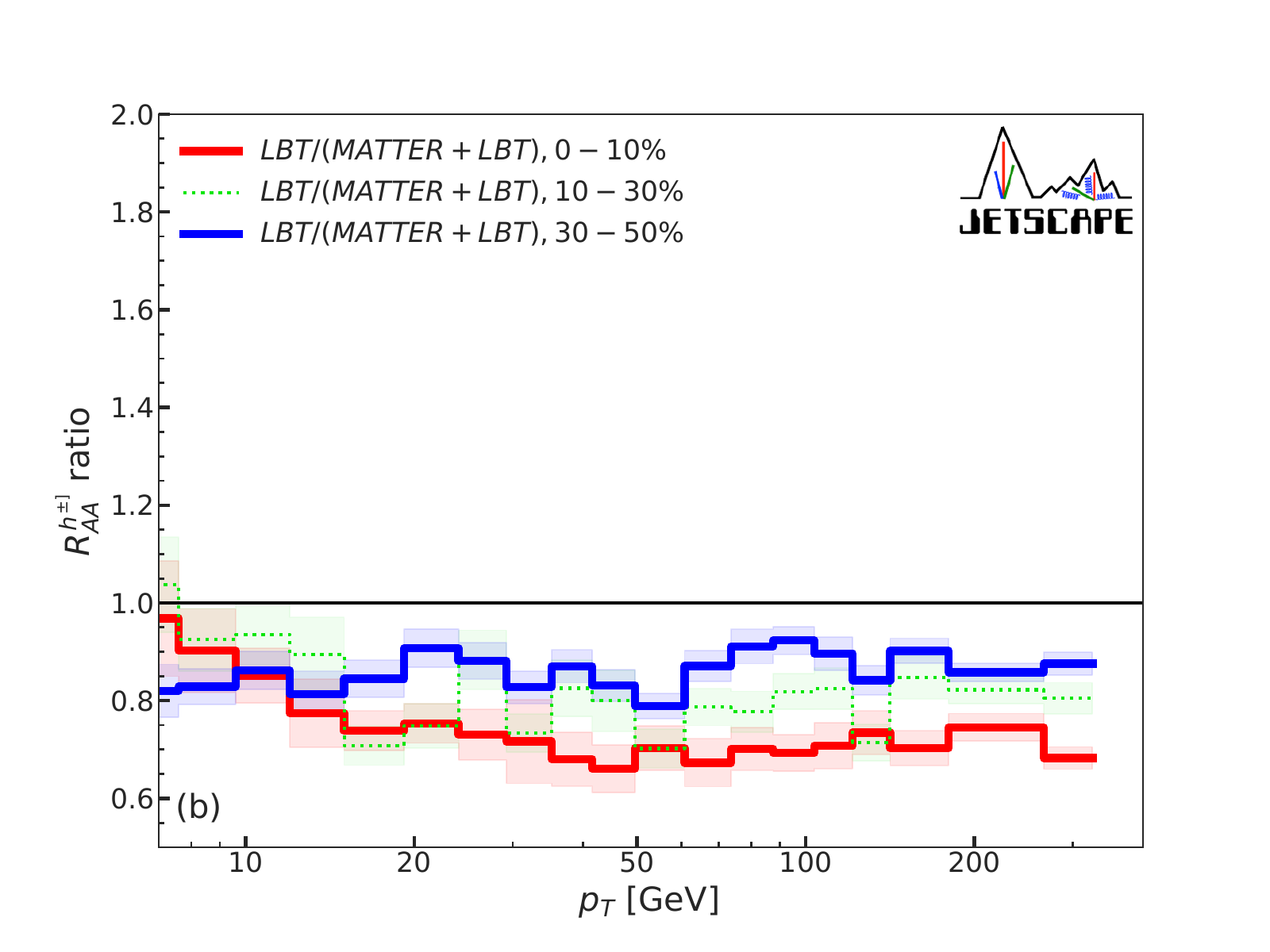}
\end{tabular}
\end{center}
\caption{(Color online) Ratio of nuclear modification factor between LBT and MATTER+LBT for D-mesons (a) and charged hadrons (b) in $\sqrt{s_{NN}}=5.02$ TeV Pb-Pb collisions at the LHC at 0-10\%, 10-30\%, 30-50\% centrality. $c_1=10, \ c_2=100$ parameters values are employed in Eq.~(\ref{eq:qhat_t}). A running $\alpha_s(\mu^2)$ is used in all calculations involving LBT.}
\label{fig:MATTER+LBT_to_LBT_ratio}
\end{figure}

The $R_{AA}$ results for the $10-30\%$ and $30-50\%$ centrality are studied in Figs.~\ref{fig:MATTER_LBT_10-30} and Fig.~\ref{fig:MATTER_LBT_30-50}. Here the ``best'' fit parameters used in the $\hat{q}(t)$ parametrization (i.e. $c_1=10, \ c_2=100$) are employed. One interesting phenomenon to notice is the reversal in order between MATTER+LBT and LBT calculation at high $p_T$ for the $D$ meson $R_{AA}$ from the most central collisions to more peripheral collisions. Two important effects contribute to this observation. First, heavy quarks are more suppressed in the MATTER phase compared to light flavor partons at high $p_T$. This can be seen from both the MATTER only simulations  in Fig.~\ref{fig:MATTER_MATTER+LBT_compare} (a) as well as in MATTER+LBT simulations, see Fig.~\ref{fig:MATTER_MATTER+LBT_compare} (b). Second, going from central to more peripheral collisions, LBT simulations seem to be more affected by the amount of time partons spend interacting with the QGP, compared to MATTER+LBT simulations. Figure~\ref{fig:MATTER+LBT_to_LBT_ratio} shows that at higher $p_T$, the ratio of $R_{AA}$ between LBT and MATTER+LBT simulations increases as the centrality increases for both D meson and charged hadrons. The virtuality dependent $\hat{q}$ reduces the in-medium contribution to MATTER evolution, making it closer to a vacuum-like (DGLAP) evolution at high $p_T$, and thus the partons spend less time in the LBT phase for MATTER+LBT simulations compared to LBT-only simulations. It is important to recall that the same $\hat{q}(t)$ is used for both light and heavy quarks throughout this work, and thus the observation that parton evolution is more vacuum-like given our parametrization for $\hat{q}$ may not necessarily hold for heavy flavors. In fact, a mass- and virtuality-dependent $\hat{q}(t,M)$ is needed to enhance future multi-scale heavy flavor evolution studies, as was mentioned when discussing Eq.~(\ref{eq:qhat_tM}). In the present work, however, as centrality increases, simulations based solely on LBT evolution are more sensitive to the reduction in QGP space-time volume compared to MATTER+LBT simulations.

Finally, note that in all our $R_{AA}$ calculations involving MATTER, the p-p baseline is using the MATTER vacuum results. The differences between PYTHIA and MATTER vacuum in p-p for charged hadrons and D mesons are found in \cite{JETSCAPE:2019udz, JETSCAPE:2020kwu}. This is essentially a comparison between PYTHIA, which generates an angular ordered shower, and MATTER which generates a virtuality ordered shower. If PYTHIA was used as the p-p baseline calculation, then the $R_{AA}$ may be further improved in some $p_T$ ranges, at the expense of calculation consistency. We choose to err on the side of a consistent calculation. 

\section{Conclusion}
\label{sec:conclusion}
This study explored how different physics entering a multi-stage description of jet partons interaction with the QGP affect both the D meson and the charged hadron $R_{AA}$. For the LBT regime, the effects of a running $\alpha_s(\mu^2)$ was studied. For the MATTER regime, we highlighted the effects of including scattering as well as considering a virtuality dependent $\hat{q}$ and found that both make a large contribution to the value of $R_{AA}$. The virtuality dependent $\hat{q}$ offers a possible explanation for the diminishing value of the interaction strength $\hat{q}/T^3$ at the LHC from previous extractions~\cite{Burke:2013yra}. However, neither of these two models alone is sufficient for describing the $R_{AA}$ at the $p_T$ range we are interested in. 

We find that the best simultaneous description of the D meson and charged hadron $R_{AA}$ requires the explicit inclusion of both the high-energy and high-virtuality regime as well as the high energy and low virtuality regime of parton energy-loss. In this work, these have been modeled using the MATTER and the LBT schemes within the JETSCAPE framework. The specific form of the $\hat{q}(t)$ parametrization is still under investigation, yet we can already state that the suppression of $\hat{q}$ at higher virtuality mostly increases $R_{AA}$ at high $p_T$. We have also explored where, in virtuality, the transition point lies between these two regimes and how changing it affects the resulting $R_{AA}$. A higher switching scale $t_s$ implies that partons will evolve longer in the LBT regime and lose more energy. While we have found that our simple exploration of the parameter space already provides a decent simultaneous description of both light and heavy flavor $R_{AA}$, a Bayesian analysis could improve the description even further. This work also shows where improved theoretical calculations are needed to better phenomenological simulations. Though a virtuality dependent $\hat{q}(t)$ does help to obtain a closer comparison between simulations to data, using solely a virtuality-dependent $\hat{q}(t)$ is not enough, both in terms of underlying physics understanding and phenomenology. Indeed, $\hat{q}(t,M)$ may explain the results found in Figs.~\ref{fig:MATTER_LBT_10-30} and Fig.~\ref{fig:MATTER_LBT_30-50}. A similar argument holds for longitudinal drag $\hat{e}$ and diffusion $\hat{e}_2$, which are know to play a part in heavy flavors physics (e.g., \cite{Cao:2017qpx}). Furthermore, the $g\to Q+\bar{Q}$ process needs to be studied using SCET that was developed in Ref.~\cite{Abir:2015hta}, given how phenomenologically important our study shows it to be (on the order of 20\%). Thus, our first phenomenological study highlights which physics of the multi-scale evolution should be improved next to better explain heavy flavor interaction in the QGP.

In the future, we plan to extend our simulation to bottom flavor and further investigate heavy flavor jet and jet substructure observables. A detailed comparison with inclusive jet observables is also of interest. The extension of our framework to such observables should provide a better constraint on $\hat{q}$. 

\section*{Acknowledgments}
\label{Ack}

This work was supported in part by the National Science Foundation (NSF) within the framework of the JETSCAPE collaboration, under grant number OAC-2004571 (CSSI:X-SCAPE). It was also supported under ACI-1550172 (Y.C. and G.R.), ACI-1550221 (R.J.F., F.G., and M.K.), ACI-1550223 (U.H., L.D., and D.L.), ACI-1550225 (S.A.B., T.D., W.F., R.W.), ACI-1550228 (J.M., B.J., P.J., X.-N.W.), and ACI-1550300 (S.C., A.K., J.L., A.M., H.M., C.N., A.S., J.P., L.S., C.Si., I.S., R.A.S. and G.V.); by PHY-1516590 and PHY-1812431 (R.J.F., M.K. and A.S.), by PHY-2012922 (C.S.); it was supported in part by NSF CSSI grant number \rm{OAC-2004601} (BAND; D.L. and U.H.); it was supported in part by the US Department of Energy, Office of Science, Office of Nuclear Physics under grant numbers \rm{DE-AC02-05CH11231} (X.-N.W.), \rm{DE-FG02-00ER41132} (D.O), \rm{DE-AC52-07NA27344} (A.A., R.A.S.), \rm{DE-SC0013460} (S.C., A.K., A.M., C.S., I.S. and C.Si.), \rm{DE-SC0021969} (C.S. and W.Z.), \rm{DE-SC0004286} (L.D., U.H. and D.L.), \rm{DE-SC0012704} (B.S.), \rm{DE-FG02-92ER40713} (J.P.) and \rm{DE-FG02-05ER41367} (T.D., W.F., J.-F.P., D.S. and S.A.B.). The work was also supported in part by the National Science Foundation of China (NSFC) under grant numbers 11935007, 11861131009 and 11890714 (Y.H. and X.-N.W.), under grant numbers 12175122 and 2021-867 (S.C.), by the Natural Sciences and Engineering Research Council of Canada (C.G., M.H., S.J., and G.V.), by the Office of the Vice President for Research (OVPR) at Wayne State University (Y.T.), by the S\~{a}o Paulo Research Foundation (FAPESP) under projects 2016/24029-6, 2017/05685-2 and 2018/24720-6 (A. L. and  M.L.), and by the University of California, Berkeley - Central China Normal University Collaboration Grant (W.K.). U.H. would like to acknowledge support by the Alexander von Humboldt Foundation through a Humboldt Research Award. C.S. acknowledges a DOE Office of Science Early Career Award. Computations were carried out on the Wayne State Grid funded by the Wayne State OVPR. The bulk medium simulations were done using resources provided by the Open Science Grid (OSG) \cite{Pordes:2007zzb, Sfiligoi:2009cct}, which is supported by the National Science Foundation award \#2030508. Data storage was provided in part by the OSIRIS project supported by the National Science Foundation under grant number OAC-1541335.

\bibliography{references}
\end{document}